\documentclass[aps,prc,amsmath,noshowpacs,superscriptaddress,nofootinbib]{revtex4}
\usepackage{graphicx,epsfig}
\newcommand{\beqy}{\begin{eqnarray}}
\newcommand{\eeqy}{\end{eqnarray}}
\newcommand{\bmlet}{\begin{subequations}}
\newcommand{\emlet}{\end{subequations}}
\newcommand{\bfdel}{\pmb{\nabla}}
\usepackage{xcolor,hyperref}

\begin{document}
\title{Pressure and chemical potentials in the inner crust of a cold neutron star within Hartree-Fock and extended Thomas-Fermi methods}
\author{N. Chamel}
\affiliation{Institut d'Astronomie et d'Astrophysique,
Universit\'e Libre de Bruxelles - CP226, 1050 Brussels,  Belgium}
\author{N.~N. Shchechilin}
\affiliation{Institut d'Astronomie et d'Astrophysique,
Universit\'e Libre de Bruxelles - CP226, 1050 Brussels,  Belgium}
\author{A.I. Chugunov}
\affiliation{Ioffe Institute, Politeknicheskaya 26, 194021  Saint Petersburg}

\date{\today} 

\begin{abstract}
Self-consistent mean-field methods with Skyrme-type effective interactions and semiclassical 
approximations, such as the Thomas-Fermi approach and its extensions are particularly well-suited 
for describing in a thermodynamically consistent way the various phases of the dense matter present 
in the interior of neutron stars. These methods have been applied to predict the composition 
of the different regions, including the inner crust constituted by nuclear clusters coexisting with free
neutrons and electrons. Because of the computational cost, the energy is typically calculated for a
few selected average baryon number densities, and the results are interpolated to obtain the pressure
numerically. However, this may introduce systematic errors in the calculations of the global 
structure of a neutron star and its dynamical evolution. 
In this paper, we show how the full equation of state can be consistently calculated 
within the same framework by deriving exact formulas for the chemical potentials and for the pressure 
that can be easily implemented in existing computer codes. These formulas are applicable to both catalyzed 
and accreted crusts. We discuss in each case the suitable conditions to impose to determine the 
composition. Numerical examples are also presented and discussed. Results from refined calculations of 
the BSk24 equation of state for the inner crust of nonaccreted neutron stars and the corresponding adiabatic 
index are provided. 
\end{abstract}

\maketitle

\section{Introduction}

A key microscopic property to describe the interior of a cold neutron star 
is the equation of state of its constituting matter, relating the pressure $P$ to the mean energy-density density $\bar \rho$ (see, e.g., 
Ref.~\cite{blaschke2018} for a review about matter in neutron stars). In the 
outer crust, the density is determined by spherical or quasispherical neutron-rich nuclei arranged in a body-centered cubic 
lattice (see, e.g., Ref.~\cite{ChamelFantina2016} and references therein) whereas the 
pressure is mainly provided by the gas of highly degenerate relativistic free electrons. For nonaccreted neutron stars under the cold catalyzed matter hypothesis~\cite{HarrisonWheeler1958}, the only microscopic inputs to 
describe this part of the crust are nuclear masses, which in principle can be experimentally measured 
in the laboratory~\cite{Wolf+2013}. The transition pressures between adjacent strata and their densities 
can be calculated analytically~\cite{Chamel20}. To a very good approximation, the pressure is given by 
that of a relativistic electron Fermi gas with corrections due to Coulomb interactions, 
exchange and charge screening (see, e.g., Ref.~\cite{cf16_exchange}). Higher-order corrections, such as 
the zero-point motion of nuclei, are negligibly small, as shown e.g. in Ref.~\cite{Pearson+11}. 

The situation is radically different in the inner region of the crust of a neutron star at mean densities above 
$\bar \rho \approx 4\times 10^{11}$ g~cm$^{-3}$, where free neutrons appear (see Ref.~\cite{Chamel_etal15_Drip} 
for further discussion about the neutron-drip transition). There neutron-rich nuclear clusters are immersed in a 
neutron liquid. Such matter cannot be reproduced in experimental facilities. The main challenge in the theoretical 
description of the inner crust is that clusters and free neutrons cannot be separately treated. The most realistic 
models rely on self-consistent mean-field methods~\cite{Bender+2003}. However, such calculations are computationally
very expensive and therefore have been so far limited to a few selected densities $\bar \rho$, as in Refs.~\cite{NegeleVautherin73,bst06,Gogelein&Muther07,Grill+11}.
The pressure is generally not calculated directly and has to be evaluated numerically a posteriori using a 
smooth fit of the total energy per nucleon. This is how the popular equation of state of Negele and 
Vautherin~\cite{NegeleVautherin73} was constructed. However, 
this may introduce systematic errors in the calculation of the global structure of a neutron star, and 
more importantly in its dynamical evolution. For instance, the smearing of density discontinuities artificially removes 
interface waves~\cite{McDermott1988}. Moreover, the fit could induce a significant loss of precision in the calculation of 
the adiabatic index $\Gamma$ characterizing the change of pressure due to variations of the average baryon number density 
$\bar n$. Because the adiabatic index enters the stellar perturbation equations, fitting errors can thus further alter the 
spectrum of oscillation modes and even the stability of the star (these issues were already raised in Chapter 10 of Ref.~\cite{htww65}). 
It is therefore desirable to calculate the pressure for a sufficient number of densities to avoid missing  
important features in the equation of state. 
For this reason, computationally much faster liquid-drop models
have been widely employed (see, e.g., Ref.~\cite{blaschke2018} and references therein). However, 
these models require some microscopic inputs, such as the surface tension, which in principle should be 
determined from mean-field calculations but are often adjusted empirically.  
Alternatively, the Thomas-Fermi (TF) method and its extensions to second or fourth order~\cite{Kirzhnits67,Grammaticos_Voros79,Grammaticos_Voros80,Brack_ea85} are 
more deeply rooted in the mean-field methods, while remaining numerically tractable and self-consistent 
(see, e.g., Ref.~\cite{ShelleyPastore2021} for comparisons). In this approach based on a semiclassical 
expansion of the Bloch density matrix in powers of $\hbar$, the local nucleon density distributions $n_q(\pmb{r})$ at position $\pmb{r}$ 
(with $q=n,p$ for neutrons, protons) are treated as basic variables instead of wavefunctions in the original mean-field method. In principle, 
$n_q(\pmb{r})$ can be found by solving the Euler-Lagrange equations arising from the variational minimization 
of the total energy of the matter element under consideration. Corrections to the energy accounting for shell effects and pairing can be added perturbatively within the Strutinsky 
integral (SI) theorem~\cite{Pearson_ea15_pairing}. Ignoring those small corrections, an analytical formula was derived in Ref.~\cite{Pearson+12} 
to calculate the pressure in this framework. In this way, numerical differentiation of the energy is avoided. This formula was applied in our 
subsequent studies, and in particular in our calculations of unified equations of state~\cite{Pearson_ea18_bsk22-26}. 
In these calculations, $n_q(\pmb{r})$ was parametrized thus reducing the variational problem to the 
minimization of the energy over a small set of parameters. However, deviations between the resulting 
nucleon densities $n_q(\pmb{r})$ and the exact solutions of the Euler-Lagrange equations introduce some 
errors in the calculations of the chemical potentials and of the pressure. Thermodynamic inconsistencies 
in the equation of state may lead to substantial errors in the global description of a neutron star (see, e.g., Refs.~\cite{Ferreira2020,Suleiman_ea21}), and could induce spurious and uncontrolled instabilities in hydrodynamical simulations.

In this paper, we analyze those errors and derive thermodynamically consistent formulas to accurately determine the pressure $P$ in the inner crust of a neutron star at any given density $\bar \rho$ directly from Hartree-Fock (HF) calculations using Skyrme-type effective interactions or from the corresponding extended TF (ETF) approximations with 
and without parametrization of the nucleon density distributions. After introducing basic thermodynamic definitions and general expressions in Section~\ref{sec:general-considerations}, both HF and ETF methods are described in Sections~\ref{sect:HF} 
and \ref{sect:ETF} respectively. We show how the pressure and chemical potentials can be calculated consistently in 
Section~\ref{sect:pressure}, where numerical applications are also presented and discussed. 
For simplicity, we do not consider pairing in this study. Its impact on the equation of state has been recently discussed elsewhere~\cite{Chamel+2024}.

\section{General considerations}
\label{sec:general-considerations}

Throughout this paper, we will focus on the region of the inner crust of a neutron star with quasi-spherical 
clusters. However, the formalism can be easily adapted to nuclear pasta. We will further assume that each layer
is a perfect crystal (presumably a body-centered or face-centered cubic lattice~\cite{Okamoto+13}) composed 
of one type of nuclear clusters and we will neglect thermal effects. In other words, we will consider matter 
in its ground state under given astrophysical conditions. 

Any matter element in the crust of a neutron star at average baryon number density $\bar n$ with $N_i$ particles 
of species $i$ ($i=e,p,n$ for electrons, protons, and neutrons) contained in a volume $V$ can be decomposed into 
$N_N$ identical cells, so-called Voronoi or Wigner-Seitz cells, 
each of which contains a single cluster at the center. In other words, $N_N$ represents the total number of clusters. The shape and volume 
\beqy 
\label{eq:cell-volume}
V_\mathrm{c}=\frac{V}{N_N} 
\eeqy 
of the Wigner-Seitz cell are 
fixed by the crystal structure (see, e.g., Ref.~\cite{Kittel1996}). In particular, the Wigner-Seitz cell of a body-centered 
cubic lattice is a truncated octahedron. 
The numbers of particles in each cell are given by 
\beqy\label{eq:total-number-particles}
N^\mathrm{(c)}_i=\frac{N_i}{N_N}\, .
\eeqy 
Each cell is electrically charge neutral and therefore contains equal numbers of 
protons and electrons: $N^\mathrm{(c)}_e=N^\mathrm{(c)}_p$. The matter element as a whole is electrically charge 
neutral and $N_e=N_p$. 

The ground-state energy $E$ of the matter element is a function of $V$, $N_N$, $N_p$ and $N_n$. 
It can be obtained from the energy $E_\mathrm{c}$ of a single cell as 
\beqy\label{eq:total-energy}
E(V,N_N,N_p,N_n)=N_N E_\mathrm{c}(V_\mathrm{c},N^\mathrm{(c)}_p,N^\mathrm{(c)}_n) \, , 
\eeqy 
where we have made use of the extensivity of the energy. The chemical potential of 
a species $i$ can thus be calculated within a single cell 
\beqy\label{eq:chemical-potential-general}
\mu_i\equiv \frac{\partial E}{\partial N_i}\biggr\vert_{V, N_N,N_{j\neq i}} =  \frac{\partial E_\mathrm{c}}{\partial N^\mathrm{(c)}_i}\Biggr\vert_{V_\mathrm{c},N^\mathrm{(c)}_{j\neq i}}\, .
\eeqy 

The pressure $P$ is defined by 
\beqy 
\label{eq:pressure-def}
P=-\frac{\partial E}{\partial V}\biggr\vert_{N_i,N_N}\, . 
\eeqy 
Inserting Eq.~\eqref{eq:total-energy} in \eqref{eq:pressure-def} using 
Eq.~\eqref{eq:total-number-particles} leads to 
\beqy\label{eq:pressure-cell}
P=-\frac{\partial E_\mathrm{c}}{\partial V_\mathrm{c}}\biggr\vert_{N^{\mathrm{(c)}}_i}\, .
\eeqy
Following Refs.~\cite{GC20_DiffEq,GKC_psi21,GC24}, the pressure can be alternatively written as 
\beqy\label{eq:pressure-general}
P=-\frac{E}{V} + \sum_i \mu_i \bar n_i +\mu_N \bar n_N \, ,
\eeqy 
where $\bar n_i=N_i/V=N^{(\mathrm{c})}_i/V_\mathrm{c}$ denotes the average particle number density of species $i$, 
$\bar n_N=1/V_\mathrm{c}$ is the average number density of clusters, and 
$\mu_N$ their `effective' chemical potential defined by 
\beqy\label{eq:chemical-potential-cluster-general} 
\mu_N\equiv \frac{\partial E}{\partial N_N}\biggr\vert_{V, N_{i}} \, .
\eeqy 
Strictly speaking, this quantity is not a true chemical potential: physically, it represents the change of energy 
associated with the creation or destruction of spatial inhomogeneities without changing the overall numbers of neutrons, protons, and electrons.\footnote{The chemical potential denoted by $\mu_N$ in 
Refs.~\cite{GC20_DiffEq,GKC_psi21,GC24} has a slightly different meaning: it represents the energy associated with the creation or destruction of spatial inhomogeneities at fixed total volume $V$ and total number of nucleons but not separate numbers of protons and neutrons. } 
From the extensivity of the energy, Eqs.~\eqref{eq:cell-volume} and \eqref{eq:total-energy}, Eq.~\eqref{eq:pressure-general} can be 
equivalently expressed as
\beqy\label{eq:pressure-general2}
P=-\frac{E_\mathrm{c}}{V_\mathrm{c}} + \sum_i \mu_i \bar n_i +\mu_N \bar n_N \, .
\eeqy
Using Eq.~\eqref{eq:cell-volume}, Eq.~\eqref{eq:chemical-potential-cluster-general} can be written as 
\beqy\label{eq:chemical-potential-cluster-cell}
\mu_N=-\frac{V_\mathrm{c}^2}{V} \frac{\partial E}{\partial V_\mathrm{c}}\biggr\vert_{V, N_{i}} \, .
\eeqy 
The derivative of the energy is explicitly given by 
\beqy\label{eq:derivative-energy-Vc}
\frac{\partial E}{\partial V_\mathrm{c}}\biggr\vert_{V, N_{i}}=\frac{V}{V_\mathrm{c}}\left( -\frac{E_\mathrm{c}}{V_\mathrm{c}}+\frac{\partial E_\mathrm{c}}{\partial V_\mathrm{c}}\biggr\vert_{N^{\mathrm{(c)}}_{i}}+\sum_i\mu_i \bar n_i\right)\, .
\eeqy 
Inserting Eq.~\eqref{eq:chemical-potential-cluster-cell} in \eqref{eq:pressure-general} using \eqref{eq:derivative-energy-Vc}, it can be easily checked that \eqref{eq:pressure-general} reduces to \eqref{eq:pressure-cell}. Substituting Eq.~\eqref{eq:derivative-energy-Vc} into \eqref{eq:chemical-potential-cluster-cell}, the chemical potential of clusters can be calculated within a single cell as 
\beqy\label{eq:chemical-potential-cluster-cell-explicit}
\mu_N= E_\mathrm{c}-V_\mathrm{c} \frac{\partial E_\mathrm{c}}{\partial V_\mathrm{c}}\biggr\vert_{N^{\mathrm{c}}_{i}}+ V_\mathrm{c} \sum_i\mu_i \bar n_i 
=P V_\mathrm{c}+E_\mathrm{c} -\sum_i \mu_i N_i^{\mathrm{(c)}}\, .
\eeqy 

The internal constitution of the inner crust of a neutron star is not a priori known. It is usually determined 
assuming matter is in equilibrium with respect to some weak and strong nuclear interactions depending on the 
astrophysical conditions. The composition is thus found by minimizing the energy $E$ for fixed volume $V$
\beqy\label{eq:energy-minimization-general}
\delta E=0=\mu_n \delta N_n + \mu_p \delta N_p+\mu_e \delta N_e +\mu_N \delta N_N\, . 
\eeqy 
Because of baryon number conservation, the variations of the numbers of nucleons are not independent but must obey 
\beqy 
\delta N_n+\delta N_p=0\, .
\eeqy 
Moreover, the condition of electric charge neutrality further requires 
\beqy 
\delta N_p=\delta N_e \, .
\eeqy 
Eq.~\eqref{eq:energy-minimization-general} thus reduces to 
\beqy\label{eq:energy-minimization}
(\mu_n-\mu_p-\mu_e) \delta N_n  +\mu_N \delta N_N=0\, . 
\eeqy 

The interior of a nonaccreted neutron star is usually determined assuming matter is 
fully ``catalyzed''~\cite{HarrisonWheeler1958}, i.e. in full thermodynamic equilibrium
with respect to all kinds of reactions. The energy is thus minimized varying all particle numbers. 
The equilibrium condition for the number of clusters (or equivalently  the volume of the cell 
from Eq.~\eqref{eq:cell-volume}) 
is given by (this condition was previously derived in Ref.~\cite{GC20_DiffEq})
\beqy\label{eq:optimum-cluster-chemical-potential}
\mu_N=0 \, , 
\eeqy
or equivalently
\beqy\label{eq:optimum-cell}
\frac{\partial E}{\partial V_\mathrm{c}}\biggr\vert_{V,N_i}=0\, .
\eeqy 
Equation~\eqref{eq:energy-minimization} 
then leads to 
\beqy\label{eq:beta-equilibrium}
\mu_n=\mu_p+\mu_e \, ,
\eeqy 
and the pressure~\eqref{eq:pressure-general2} reduces to 
\beqy
\label{eq:pressure-catalyzed}
P=\bar n\mu_n - \frac{E_\mathrm{c}}{V_\mathrm{c}} \, ,
\eeqy 
where $\bar n=\bar n_n+\bar n_p$. 

In the crust of accreted neutron stars, the number of clusters $N_N$  (therefore also $V_\mathrm{c}$ from Eq.~\eqref{eq:cell-volume}) is fixed in the absence of pycnonuclear fusion reactions. 
In this case, Eq.~\eqref{eq:energy-minimization} also leads to the same condition~\eqref{eq:beta-equilibrium}
but the pressure~\eqref{eq:pressure-general2} is now given by 
\beqy
\label{eq:pressure-accreted}
P=\bar n\mu_n + \bar n_N \mu_N - \frac{E_\mathrm{c}}{V_\mathrm{c}} \, . 
\eeqy 
This coincides with the expression previously derived in Refs.~\cite{GC20_DiffEq,GKC_psi21,GC24}. 

If $N_p^{\mathrm{(c)}}$ is fixed but $N_n^{\mathrm{(c)}}$ and $N_N$ are allowed to vary, the equilibrium condition~\eqref{eq:energy-minimization}  becomes
\beqy\label{eq:mu_N_fixZ}
\bar n \mu_n =\sum_i \bar n_i \mu_i + \bar n_N \mu_N \, ,
\eeqy 
while the pressure~\eqref{eq:pressure-general2} coincides with Eq.~\eqref{eq:pressure-catalyzed}.

Although Eq.~\eqref{eq:pressure-general2} is seemingly more practical than Eq.~\eqref{eq:pressure-cell} by avoiding the numerical evaluation of the partial derivative of the energy with respect to the volume of the cell, its reliability relies on the consistent and accurate determination of the chemical potentials \eqref{eq:chemical-potential-general}. 

Explicit expressions for the chemical potentials and the pressure within the HF theory and the ETF approximation will be derived in the following sections.

\section{Hartree-Fock theory of neutron-star crust}
\label{sect:HF}

\subsection{Hartree-Fock equations}

In the inner crust of neutron star, electron charge-screening effects are sufficiently small~\cite{Maruyama2005} that the local number density of electrons $n_e(\pmb{r})$ at position $\pmb{r}$ inside the Wigner-Seitz cell deviates only slightly from the average value 
\beqy 
\bar n_e = \frac{1}{V_\mathrm{c}}\int{\rm d}^3\pmb{r}\, n_e(\pmb{r})=\frac{N^{\mathrm{(c)}}_e}{V_\mathrm{c}}\, .
\eeqy 
To a very good approximation, electrons can be described semiclassically as a degenerate relativistic Fermi gas with small corrections due to Coulomb interactions  (see, e.g., Ref.~\cite{Salpeter1961}).  Nucleons are described in the 
Hartree-Fock (HF) theory with zero-range effective interactions of the Skyrme type~\cite{Bender+2003}. The nucleon densities at position $\pmb{r}$ will 
be denoted by $n_q(\pmb{r})$ with 
\beqy 
N^{(\mathrm{c})}_q=\int{\rm d}^3\pmb{r}\, n_q(\pmb{r})\, .
\eeqy

Adopting the Kohn-Sham approximation for Coulomb exchange~\cite{KohnSham1965},  the energy of the cell
can be written as 
the integral of a semilocal energy density functional 
\beqy\label{eq:E_HF}
E_{\rm HF} = \int{\rm d}^3\pmb{r}\, \mathcal{E}_{\rm HF}\Big[n_q(\pmb{r}),\tau_q(\pmb{r}), 
\pmb{J_q}(\pmb{r}),n_e(\pmb{r})\Big] \, ,
\eeqy 
where $\tau_q(\pmb{r})$ is the nucleon kinetic-energy density (in units of $\hbar^2/2m_q$ with $\hbar$ the Planck-Dirac constant and 
$m_q$ the nucleon mass), and 
$\pmb{J_q}(\pmb{r})$ is the nucleon spin current density. This functional can be decomposed into three terms: 
\beqy \label{eq:energy-contributions}
\mathcal{E}_{\rm HF}\Big[n_q(\pmb{r}),\tau_q(\pmb{r}), 
\pmb{J_q}(\pmb{r}),n_e(\pmb{r})\Big]=\mathcal{E}_e\Big[n_e(\pmb{r})\Big] + \mathcal{E}_{\rm nuc}\Big[n_q(\pmb{r}),\tau_q(\pmb{r}), 
\pmb{J_q}(\pmb{r})\Big] +  \mathcal{E}_{\rm Coul}\Big[n_p(\pmb{r}),n_e(\pmb{r})\Big] \, .
\eeqy 
The first term accounts for the local kinetic energy density of an ideal relativistic electron Fermi gas of density $n_e(\pmb{r})$, the second term is a purely nuclear 
contribution, and the last term accounts for Coulomb interactions. The Coulomb part consists of a direct term 
\beqy \label{eq:direct-Coulomb-energy}
{\mathcal E}_{\rm Coul, dir}(\pmb{r})=\frac{e}{2} n_{\rm ch}(\pmb{r}) U_{\rm Coul}(\pmb{r}) \, ,
\eeqy 
where $U_{\rm Coul}(\pmb{r})$ is the solution of Poisson's equation with the charge density $n_{\rm ch}(\pmb{r})=n_p(\pmb{r})-n_e(\pmb{r})$, 
\beqy 
\nabla^2 U_{\rm Coul}(\pmb{r}) = -4 \pi e n_{\rm ch}(\pmb{r}) \, ,
\eeqy 
and an exchange term given by 
\beqy 
{\mathcal E}_{\rm Coul, exch}(\pmb{r})=-\frac{3 e^2}{4}\left(\frac{3}{\pi}\right)^{1/3}\left( x n_p(\pmb{r})^{4/3}-\frac{1}{2} n_e(\pmb{r})^{4/3}\right) \, .
\eeqy 
The parameter $x$ is generally equal to 1, except for some of the Brussels-Montreal functionals adopted in our previous studies 
and for which $x=0$. 

Introducing the nucleon density matrix in coordinate space $n_q(\pmb{r}, \sigma; \pmb{r^\prime}, \sigma^\prime)$ (with $\pmb{r}, \pmb{r^\prime}$ the spatial 
coordinates and $\sigma,\sigma^\prime=\pm1$ the spin coordinates), the local nucleon densities and currents are defined by 
\beqy\label{eq:rhoq}
n_q(\pmb{r}) = \sum_{\sigma=\pm 1}n_q(\pmb{r}, \sigma; \pmb{r}, \sigma)
\, ,
\eeqy
\beqy\label{eq:tauq}
\tau_q(\pmb{r}) = \sum_{\sigma=\pm 1}\int\,{\rm d}^3\pmb{r^\prime}\,\delta(\pmb{r}-\pmb{r^\prime}) \bfdel\cdot\bfdel^\prime
n_q(\pmb{r}, \sigma; \pmb{r^\prime}, \sigma) \, ,
\eeqy
\beqy\label{eq:Jq}
\pmb{J_q}(\pmb{r}) = -{\rm i}\sum_{\sigma,\sigma^\prime=\pm1}\int\,{\rm d}^3\pmb{r^\prime}\,\delta(\pmb{r}-\pmb{r^\prime}) 
\bfdel n_q(\pmb{r}, \sigma; \pmb{r^\prime},
\sigma^\prime) \times \mbox{\boldmath$\sigma$}_{\sigma^\prime \sigma}   \nonumber \\
={\rm i}\sum_{\sigma,\sigma^\prime=\pm1}\int\,{\rm d}^3\pmb{r^\prime}\,\delta(\pmb{r}-\pmb{r^\prime}) 
\bfdel^\prime n_q(\pmb{r}, \sigma; \pmb{r^\prime},
\sigma^\prime) \times \mbox{\boldmath$\sigma$}_{\sigma^\prime \sigma} \, ,
\eeqy
where $\mbox{\boldmath$\sigma$}_{\sigma\sigma^\prime}$ denotes the Pauli spin 
matrices. In turn, the nucleon density matrix is defined by 
\beqy
n_q(\pmb{r}, \sigma; \pmb{r^\prime}, \sigma^\prime) = \langle\Psi|c_q(\pmb{r^\prime}\sigma^\prime)^\dagger c_q(\pmb{r}\sigma)|\Psi\rangle \, ,
\eeqy
where $|\Psi\rangle$ is the ground-state many-body wave function of $N_q$ nucleons,  $c_q(\pmb{r}\sigma)^\dagger$ and 
$c_q(\pmb{r}\sigma)$ are the creation and destruction operators for nucleons of charge type $q$ 
at position $\pmb{r}$ with spin projection $\sigma \hbar$. In the HF theory, $|\Psi\rangle$ is written as a Slater determinant. 

The state of the cell is found   by minimizing the energy $E_{\rm HF}$ with fixed numbers $N^{(\mathrm{c})}_q$ of nucleons and number $N^{(\mathrm{c})}_e$ of electrons in the cell,
or equivalently by minimizing the grand potential $E_{\rm HF} - \sum_q \mu_q N^{(\mathrm{c})}_q -\mu_e N^{(\mathrm{c})}_e$, where $\mu_q$ and $\mu_e$ are Lagrange multipliers. Variations of this potential, 
\beqy\label{eq:delta_E_HF0}
\delta E_{\rm HF} - \sum_q \mu_q \delta N^{(\mathrm{c})}_q -\mu_e \delta N^{(\mathrm{c})}_e=0 \, ,
\eeqy 
leads to the Euler-Lagrange equation for the electrons 
\beqy\label{eq:Euler-Lagrange-electrons-HF}
\mu_e= \frac{\delta E_{\rm HF}}{\delta n_e(\pmb{r})} \, ,
\eeqy
and to the self-consistent HF equations for the nucleons
\beqy\label{eq:HF}
\sum_{\sigma^\prime} h_q(\pmb{r})_{\sigma\sigma^\prime} \varphi^{(q)}_{k}(\pmb{r}, \sigma^\prime) = \epsilon^{(q)}_k \varphi^{(q)}_{k}(\pmb{r}, \sigma)\, .
\eeqy
Here the single-particle wave functions $\varphi^{(q)}_{k}(\pmb{r}, \sigma)$ are characterized by the set of quantum numbers $k$; $\epsilon^{(q)}_k$ are Lagrange multipliers introduced to enforce the normalization of the wave functions $\varphi^{(q)}_{k}(\pmb{r}, \sigma)$. The nucleon density matrix is then given by 
\beqy\label{eq:HFrho}
n_q(\pmb{r}, \sigma; \pmb{r^\prime}, \sigma^\prime) = \sum_{k(q)} n^{(q)}_k \varphi^{(q)}_{k}(\pmb{r}, \sigma) \varphi^{(q)}_{k}(\pmb{r^\prime}, \sigma^\prime)^* \, ,
\eeqy
where $n^{(q)}_k$ is the occupation factor of the single particle state $k$ given by 
the Heaviside unit-step function $H$
\beqy
n^{(q)}_k = H(\mu_q-\epsilon^{(q)}_k)\, .
\eeqy
The single-particle Hamiltonian is 
\beqy\label{eq:hq}
h_q(\pmb{r})_{\sigma\sigma^\prime} \equiv -\pmb{\nabla}\cdot \frac{\hbar^2}{2 m_q^\oplus(\pmb{r})} \pmb{\nabla}\delta_{\sigma\sigma^\prime}
+  U_q(\pmb{r}) \delta_{\sigma\sigma^\prime}
-{\rm i}
\pmb{W_q}(\pmb{r})\cdot \pmb{\nabla} \times \pmb{\sigma}_{\sigma\sigma^\prime} \, ,
\eeqy
where 
\beqy\label{eq:fields}
\frac{\hbar^2}{2 m_q^\oplus(\pmb{r})} = 
\frac{\partial \mathcal{E}_{\rm HF}(\pmb{r})}{\partial\tau_q(\pmb{r})}\, , 
\hskip0.5cm 
U_q(\pmb{r})=\frac{\delta E_{\rm HF}}
{\delta n_q(\pmb{r})}\, , \hskip0.5cm
\pmb{W_q}(\pmb{r})=\frac{\partial \mathcal{E}_{\rm HF}(\pmb{r})}
{\partial
\pmb{J_q}(\pmb{r})}  \, .
\eeqy
In the above expression, we have introduced the functional derivative (see Appendix~\ref{app:functional-derivative})
\beqy\label{eq:Uq}
\frac{\delta E_{\rm HF}}{\delta n_q( \pmb{r})}\equiv\frac{\partial\,\mathcal{E}_{\rm HF}(\pmb{r})}{\partial\,n_q( \pmb{r})}-\pmb{\nabla}\cdot\frac{\partial\,\mathcal{E}_{\rm HF}(\pmb{r})}{\partial\,\pmb{\nabla}n_q(\pmb{r})}\, ,
\eeqy
recalling that the Skyrme energy density functional involves only the first derivatives of the nucleon densities
(after suitable integrations by parts).
Note that with our definition, the proton mean field potential $U_p(\pmb{r})$ contains direct and exchange Coulomb contributions. 
The self-consistency of the HF equations~\eqref{eq:HF} arises from the following expressions for the densities and currents: 
\beqy 
n_q(\pmb{r}) =  \sum_{k(q)}n^{(q)}_k \sum_{\sigma}\vert\varphi^{(q)}_{k}(\pmb{r}, \sigma)\vert^2 \, ,
\eeqy 
\beqy 
\tau_q(\pmb{r}) =  \sum_{k(q)}n^{(q)}_k \sum_{\sigma} \vert\pmb{\nabla} \varphi^{(q)}_{k}(\pmb{r}, \sigma)\vert^2 \, ,
\eeqy 
\beqy 
\pmb{J_q}(\pmb{r})={\rm i} \sum_{k(q)}n^{(q)}_k \sum_{\sigma,\sigma^\prime} \varphi^{(q)}_{k}(\pmb{r}, \sigma)\pmb{\nabla}\varphi^{(q)}_{k}(\pmb{r}, \sigma^\prime)^* \times \pmb{\sigma}_{\sigma^\prime\sigma}\, .
\eeqy 

By symmetry, the Euler-Lagrange equations~\eqref{eq:Euler-Lagrange-electrons-HF} for electrons must be solved with periodic boundary conditions 
and the HF equations~\eqref{eq:HF} for nucleons with Floquet-Bloch boundary conditions 
(see, e.g., Ref.~\cite{Kittel1996}). Once the variational solution has been obtained, it follows from Eq.~\eqref{eq:delta_E_HF0} that the Lagrange multipliers $\mu_q$ and $\mu_e$
\beqy 
\mu_q=\frac{\partial E_{\rm HF}}{\partial N^{(\mathrm{c})}_q}\biggr\vert_{N^{(\mathrm{c})}_e,N^{(\mathrm{c})}_{q'\neq q}}\, ,
\eeqy 
\beqy 
\mu_e=\frac{\partial E_{\rm HF}}{\partial N^{(\mathrm{c})}_e}\biggr\vert_{N^{(\mathrm{c})}_q}\, 
\eeqy 
coincide with the nucleon and electron chemical potentials respectively, see Eq.~\eqref{eq:chemical-potential-general}. For catalyzed matter, such HF calculations must be repeated varying the composition, the crystal structure and the volume of the cell to find the absolute ground state.

At high enough densities, the crust dissolves into a homogeneous mixture of neutrons, protons and electrons, this 
marks the transition to the outer core. 
This transition can be consistently described within the HF theory. 
In this case, the HF equations~\eqref{eq:HF} have to be solved inside the entire volume $V$ of the matter element with Born-von K\'arm\'an periodic boundary conditions and with $n_e(\pmb{r})=\bar n_e$. The nucleon single-particle wave functions 
reduce to plane waves 
\beqy
\varphi^{(q)}_{k}(\pmb{r}, \sigma)=\frac{1}{\sqrt{V}}\exp({\rm i} \pmb{k}\cdot\pmb{r}) \chi^{(q)}(\sigma)\, , 
\eeqy
where $\chi^{(q)}(\sigma)$ denotes Pauli spinor. Here the volume $V$ of the matter element is chosen to be large enough for the components of the wave vector 
$\pmb{k}$ to be treated as quasi-continuous. The nucleon single-particle energies can be readily obtained from Eq.~(\ref{eq:HF})~:
\beqy
\epsilon^{(q)}_{\pmb{k}}=\frac{\hbar^2 k^2}{2 m_q^\oplus} + U_q\, .
\eeqy
The nucleon chemical potentials then coincide with the Fermi energies 
\beqy\label{eq:muq_HF}
\mu_q = \frac{\hbar^2 k_{{\rm F}q}^2}{2 m_q^\oplus} + U_q-x\delta_{q,p} e^2\left(\frac{3}{\pi}\right)^{1/3} n_p^{1/3}\, , \hskip 0.5cm k_{{\rm F} q}=(3\pi^2 n_q)^{1/3}\, ,
\eeqy
\beqy\label{eq:mue_hom}
\mu_e = \frac{d \mathcal{E}_e}{d n_e} + \frac{e^2}{2} \left(\frac{3}{\pi}\right)^{1/3}n_e^{1/3} \, .
\eeqy
In this region of the star,  the pressure can be calculated analytically 
(see, e.g., Ref.~\cite{Pearson_ea18_bsk22-26}). 

\subsection{Wigner-Seitz approximation}

Solving the coupled HF Eqs.~\eqref{eq:HF} in the inner crust for both neutrons and protons imposing Floquet-Bloch boundary conditions 
is very challenging. A popular approximation introduced by Wigner and Seitz~\cite{WignerSeitz33} in the context of electrons in solids and first implemented 
in neutron star-crusts by Negele and Vautherin~\cite{NegeleVautherin73}, consists in replacing the exact Wigner-Seitz cell by a spherical cell of radius $R$. 
The nucleon densities and currents inside this cell are further assumed to be spherically symmetric. 

The nucleon wave function (assuming closed shells) can then be decomposed as~\cite{Vautherin_Brink1972}
\beqy 
\varphi^{(q)}_{\alpha}(\pmb{r}, \sigma) =\mathcal{R}^{(q)}_{\alpha}(r) \mathcal{Y}_{\ell j m}(\theta, \phi, \sigma)\, ,
\eeqy 
\beqy 
\mathcal{Y}_{\ell j m}(\theta,\phi, \sigma)=\sum_{m_\ell,m_s}\langle \ell \frac{1}{2} m_\ell m_s \vert j m\rangle Y_{\ell m_\ell}(\theta,\phi)\chi^{(q)}_{m_s}(\sigma) \, ,
\eeqy 
where $\alpha$ stands for the set of quantum numbers: the principal quantum number, the orbital angular momentum $\ell$, and total angular momentum $j$. The nucleon densities and currents read~\cite{Vautherin_Brink1972}
\beqy 
n_q(r) = \frac{1}{4\pi} \sum_{\alpha(q)}g^{(q)}_\alpha \mathcal{R}^{(q)}_{\alpha}(r)^2 \, ,
\eeqy 
\beqy 
\tau_q(r) = \frac{1}{4\pi} \sum_{\alpha(q)}g^{(q)}_\alpha \Biggl[\left(\frac{d\mathcal{R}^{(q)}_{\alpha}}{dr}\right)^2+\frac{\ell_\alpha(\ell_\alpha+1)}{r^2} \mathcal{R}^{(q)}_{\alpha}(r)^2\Biggr] \, ,
\eeqy 
\beqy 
J_q(r) = \frac{1}{4\pi r} \sum_{\alpha(q)}g^{(q)}_\alpha  \left[j_\alpha(j_\alpha+1)-\ell_\alpha(\ell_\alpha+1)-\frac{3}{4}\right]\mathcal{R}^{(q)}_{\alpha}(r)^2 \, ,
\eeqy 
where $g^{(q)}_\alpha\equiv n^{(q)}_\alpha (2j_\alpha+1)$. 

The HF equations~\eqref{eq:HF} thus greatly simplify since they reduce to ordinary differential equations~\cite{Vautherin_Brink1972} 
\beqy \label{eq:HF-WS}
&\Biggl\{-\dfrac{1}{r^2}\dfrac{d}{dr} r^2 
\dfrac{\hbar^2}{2 m_q^\oplus(r)} \dfrac{d}{dr} +\dfrac{\hbar^2 \ell_\alpha(\ell_\alpha+1)}{2 m_q^\oplus(r) r^2}  + U_q(r)+ \left[j_\alpha(j_\alpha+1)-\ell_\alpha(\ell_\alpha+1)-\dfrac{3}{4}\right]\dfrac{W_q(r)}{r}\Biggr\}  \mathcal{R}^{(q)}_{\alpha} (r) \notag \\ 
&= \epsilon^{(q)}_\alpha \mathcal{R}^{(q)}_{\alpha} (r) \, .
\eeqy 
These equations have been solved with different prescriptions for the boundary conditions. Although protons are typically found to be localized near the center of the cell and their wavefunctions are therefore exponentially suppressed at the cell edge (except possibly in the deepest layers of the crust where some protons are loosely bound or might even be unbound), the proton states can still be sensitive to the choice of boundary conditions through the neutron contributions to the proton single-particle
Hamiltonian and the self-consistency of the HF Eqs.~\eqref{eq:HF-WS}. 
Two types of boundary conditions have been considered: 
the Dirichlet boundary conditions 
\beqy\label{eq:Dirichlet}
\mathcal{R}^{(q)}_{\alpha} (R)=0 \, ,
\eeqy 
or the Neumann boundary conditions 
\beqy \label{eq:Neumann}
\frac{d\mathcal{R}^{(q)}_{\alpha}}{dr}\biggr\vert_{r=R}=0 \, . 
\eeqy 
The first choice, adopted for instance in Refs.~\cite{Pizzochero+2002,Barranco+2010,Baroni+2010}, leads to a neutron depletion in the interstitial region between clusters, and this is clearly nonphysical. Band-structure calculations showed that unbound neutrons 
are essentially uniformly distributed sufficiently far from clusters~\cite{Chamel+07}. 
The second choice, proposed by Wigner and Seitz~\cite{WignerSeitz33}, may seem closer to the 
Floquet-Bloch boundary conditions. However, this translates into a spurious neutron accumulation in 
the interstitial region between clusters~\cite{Montani+04}. It should be stressed that the large underlying 
oscillations of the wavefunctions contaminate not only the densities $n_q(r)$ but also the kinetic densities 
$\tau_q(r)$ and spin currents $J_q(r)$. To avoid large density fluctuations, Negele and Vautherin~\cite{NegeleVautherin73} 
adopted instead mixed Dirichlet-Neumann boundary conditions, namely by requiring Eq.~\eqref{eq:Dirichlet} 
for even $\ell$ and Eq.~\eqref{eq:Neumann} for odd $\ell$. Because density fluctuations were still present 
(though their amplitude was reduced), Negele and Vautherin further replaced $n_n(r)$ and $\tau_n(r)$ (neutron 
spin-orbit coupling was neglected) in the vicinity of the cell edge by their average values in the interstitial 
region at each iteration of the HF calculation. This prescription, however, introduces some inconsistencies in 
the HF results. As discussed in Ref.~\cite{bst06}, imposing Eq.~\eqref{eq:Dirichlet} for odd $\ell$ and 
Eq.~\eqref{eq:Neumann} for even $\ell$ is as physically motivated as the conditions of Negele and Vautherin. 
Adopting one or the other mixed boundary conditions leads to errors in the HF energy that are significant enough 
to change the equilibrium composition. The later study of Ref.~\cite{Pastore+2011} showed that the distributions
of neutrons and protons within the whole cell are different even if the overall composition and the cell size are 
the same. These kinds of errors can hardly be removed due to the highly non-linearity of the HF equations (but see, e.g., 
Ref.~\cite{Pastore+2017}). They arise because of the Wigner-Seitz approximation and would therefore entirely vanish 
if the HF equations were solved in the exact Wigner-Seitz cell imposing Floquet-Bloch boundary conditions. However, 
such calculations are computationally very demanding. Alternatively, these errors can be circumvented without abandoning 
the Wigner-Seitz approximation by following the extended Thomas-Fermi approach.

\section{Extended Thomas-Fermi approximation}
\label{sect:ETF}

\subsection{Euler-Lagrange equations}
\label{sect:ETF_EL}

An approximate estimate for the energy $E_{\rm HF}$ 
can be obtained using the ETF 
method~\cite{Brack_ea85}, in which the kinetic-energy densities and the spin-current densities are expressed 
in terms of the densities and their gradients, which now become the basic variables, rather 
than the single-particle wave functions. Below, we consider the fourth-order expansion. 
The ETF energy of the cell can thus be written as a functional of the densities
\beqy\label{eq:E_ETF}
E_{\rm ETF} = \int{\rm d}^3\pmb{r}\, \mathcal{E}_{\rm ETF}\Big[n_q(\pmb{r}),n_e(\pmb{r})\Big]\, .
\eeqy 
The nucleon densities $n_q(\pmb{r})$ and electron densities $n_e(\pmb{r})$ are obtained by minimising the ETF energy with fixed numbers $N^{(\mathrm{c})}_q$ of nucleons and number $N^{(\mathrm{c})}_e$ of electrons, or equivalently the associated grand potential~:
\beqy\label{eq:delta_E_ETF0}
\delta E_{\rm ETF} - \sum_q \mu_q \delta N^{(\mathrm{c})}_q -\mu_e \delta N^{(\mathrm{c})}_e=0 \, .
\eeqy 
The Lagrange multipliers $\mu_q$ and $\mu_e$ then coincide with the nucleon and electron chemical potentials respectively since we have 
(compare with Eq.~\eqref{eq:chemical-potential-general}) 
\beqy\label{eq:muq_def} 
\mu_q=\frac{\partial E_{\rm ETF}}{\partial N^{(\mathrm{c})}_q}\biggr\vert_{N^{(\mathrm{c})}_e,N^{(\mathrm{c})}_{q'\neq q}}\, ,
\eeqy 
\beqy 
\mu_e=\frac{\partial E_{\rm ETF}}{\partial N^{(\mathrm{c})}_e}\biggr\vert_{N^{(\mathrm{c})}_q}\, .
\eeqy 
From the functional form of the ETF energy, Eq.~\eqref{eq:delta_E_ETF0} can be explicitly written as 
\beqy\label{eq:delta_E_ETF}
\sum_q\int{\rm d}^3\pmb{r}\,\Biggl[ \frac{\delta E_{\rm ETF}}{\delta n_q(\pmb{r})}  -  \mu_q\Biggr]\delta n_q(\pmb{r}) +\int{\rm d}^3\pmb{r}\,\Biggl[ \frac{\delta E_{\rm ETF}}{\delta n_e(\pmb{r})}  -  \mu_e\Biggr]\delta n_e(\pmb{r})=0 \, .
\eeqy 
Since the variations $\delta n_q(\pmb{r})$ and $\delta n_e(\pmb{r})$ are \emph{arbitrary} (apart from the 
condition that they vanish at the fixed boundary of the cell), the variations $\delta E_{\rm ETF}$ of the ETF 
energy thus lead to the Euler-Lagrange equations
(see Appendix~\ref{app:functional-derivative})
\beqy\label{eq:Euler-Lagrange}
\mu_q =\frac{\delta E_{\rm ETF}}{\delta n_q(\pmb{r})} = \frac{\partial \mathcal{E}_{\rm ETF}(\pmb{r})}{\partial n_q(\pmb{r})} - \pmb{\nabla}\cdot \frac{\partial \mathcal{E}_{\rm ETF}(\pmb{r})}{\partial \pmb{\nabla}n_q(\pmb{r})} + \nabla^2 \frac{\partial \mathcal{E}_{\rm ETF}(\pmb{r})}{\partial \nabla^2 n_q(\pmb{r})}
- \pmb{\nabla}\cdot \nabla^2 \frac{\partial \mathcal{E}_{\rm ETF}(\pmb{r})}{\partial \pmb{\nabla} \nabla^2 n_q(\pmb{r})} 
+ \nabla^4 \frac{\partial \mathcal{E}_{\rm ETF}(\pmb{r})}{\partial \nabla^4 n_q(\pmb{r})}
\, ,
\eeqy
\beqy\label{eq:Euler-Lagrange-electrons}
\mu_e= \frac{\delta E_{\rm ETF}}{\delta n_e(\pmb{r})} \, .
\eeqy
These equations must hold at \emph{any} point $\pmb{r}$ inside the volume of the cell and must be solved with periodic boundary conditions. 

It is worth recalling that the TF theory of ordinary solids was shown to be exact in the limit when the number of electrons tends to infinity~\cite{LiebSimon1973}. The ETF approach is therefore expected to be particularly well-suited for describing unbound neutrons in the inner crust of a neutron star since typically $N^{(\mathrm{c})}_n\sim 10^2-10^3\gg N^{(\mathrm{c})}_p\sim 40$ (see, e.g., Ref.~\cite{Pearson_ea18_bsk22-26}). 

In the limit of homogeneous matter with given densities $n_q$, as in the outer core of neutron stars, the ETF method yields the same energy as HF. Moreover, it can be easily seen that in this case Eq.~(\ref{eq:Euler-Lagrange}) coincides exactly with the HF expression~(\ref{eq:muq_HF}) for the chemical potentials. The pressure is also the same as in the HF theory.

\subsection{Wigner-Seitz approximation}
\label{Sec:ETF_WS}

As for HF calculations of the inner crust of a neutron star, the spherical Wigner-Seitz cell approximation has been widely adopted within the TF approach and its extensions. In this case, 
the densities reduce to functions of the radial coordinate $r$ only. 
The periodic boundary conditions are replaced by 
\beqy\label{eq:WS}
\frac{dn_q}{dr}\biggr\vert_{r=0} = 0\, , \hskip0.5cm \frac{dn_q}{dr}\biggr\vert_{r=R} = 0\, ,
\eeqy 
and similarly for electrons. 

At the lowest order in the ETF expansion, we have $\displaystyle \tau_p(r)= \frac{3}{5}(3\pi^2)^{2/3} n_p(r)^{5/3}$. 
The proton chemical potential evaluated at the cell edge can be expressed as 
\beqy
\mu_p = \frac{\hbar^2}{2 m_p^\oplus(R)} \frac{\partial \tau_p(R)}{\partial n_p(R)} + U_p(R) \, .
\eeqy
Let us assume that protons are tightly bound in clusters centered around $r=0$ so that $n_p(R) \approx 0$ therefore  
\beqy
 \frac{\partial \tau_p(R)}{\partial n_p(R)}=[3\pi^2 n_p(R)]^{2/3}\approx 0\, ,
\eeqy
and consequently $\mu_p\approx U_p(R)$ independently of proton binding strength ($U_p(R)$ does not vanish due to the presence of unbound neutrons). This shows that the simple TF approximation fails to describe bound particles (as is well-known for electrons in ordinary atoms and in the envelope of neutron stars~\cite{ShapiroTeukolsky}) and must be extended to obtain more realistic estimates of the proton chemical potential.

\subsection{Parametrized particle distributions}
\label{sect:ETF_par}

Solving the Euler-Lagrange Eqs.~(\ref{eq:Euler-Lagrange}) still remains computationally expensive. Moreover, these nonlinear equations are prone to numerical instabilities when second and fourth-order terms are included~\cite{Brack_ea85}. A much simpler approach consists in minimizing the 
ETF energy~(\ref{eq:E_ETF}) using parametrized particle density distributions, which can be quite generally written as 
\beqy\label{eq:parametrized-profiles1}
n_q(r) = n_{Bq} + n_{\Lambda q}f_q(r,\pmb{x_q},R)  \, ,
\eeqy
in which $n_{Bq}$ is a constant background term whereas the second term accounts for the presence of a cluster associated with a density excess $n_{\Lambda q}$
centered around $r=0$; the shape of the cluster is described by the dimensionless function $f_q(r,\pmb{x_q},R)$ where $\pmb{x_q}$ denotes a set of geometric parameters. Note that this function must also depend explicitly on $R$ to fulfill the boundary condition~\eqref{eq:WS}. 

Because of the restrictions imposed on the nucleon density distributions, the resulting ground-state energy $\tilde{E}_{\rm ETF}$ only provides an upper bound for 
the exact ETF energy $E_{\rm ETF}$. Moreover, the parametrized profiles~\eqref{eq:parametrized-profiles1} do not satisfy the Euler-Lagrange Eqs.~\eqref{eq:Euler-Lagrange}, however, an analog of these equations can be derived considering that the variations $\delta n_q(\pmb{r})$ 
of the nucleon densities are not completely arbitrary but are now restricted to~:
\beqy\label{eq:variation_param-profiles}
\delta n_q(r) = \frac{\partial n_q(r)}{\partial n_{Bq}}\delta n_{Bq} + \frac{\partial n_q(r)}{\partial \pmb{x_q}}\cdot \delta \pmb{x_q} 
+ \frac{\partial n_q(r)}{\partial n_{\Lambda q}}\delta n_{\Lambda q}\, .
\eeqy
Inserting these variations in Eq.~(\ref{eq:delta_E_ETF}) for arbitrary $\delta n_{Bq}$, $\delta n_{\Lambda q}$, and $\delta\pmb{x_q}$ thus leads to 
\beqy\label{eq:delta_E_ETF_param1}
\int{\rm d}^3\pmb{r}\,\frac{\partial n_q(\pmb{r})}{\partial n_{Bq}} \Bigl[ \frac{\delta E_{\rm ETF}}{\delta n_q(\pmb{r})}  -  \tilde{\mu}_q\Bigr]=0 \, ,
\eeqy 
\beqy\label{eq:delta_E_ETF_param2}
\int{\rm d}^3\pmb{r}\,\frac{\partial n_q(\pmb{r})}{\partial \pmb{x_{q}}} \Bigl[ \frac{\delta E_{\rm ETF}}{\delta n_q(\pmb{r})}  -  \tilde{\mu}_q\Bigr]=0 \, ,
\eeqy 
\beqy\label{eq:delta_E_ETF_param4}
\int{\rm d}^3\pmb{r}\,\frac{\partial n_q(\pmb{r})}{\partial n_{\Lambda q}} \Bigl[ \frac{\delta E_{\rm ETF}}{\delta n_q(\pmb{r})}  -  \tilde{\mu}_q\Bigr]=0 \, .
\eeqy 
The partial differential Euler-Lagrange Eqs.~(\ref{eq:Euler-Lagrange}) in the original 
ETF method have thus been replaced by a set of nonlinear coupled equations for the parameters $n_{Bq}$, $n_{\Lambda q}$, $\pmb{x_q}$. 
Using
Eq.~(\ref{eq:parametrized-profiles1}), Eq.~(\ref{eq:delta_E_ETF_param1}) yields 
\beqy\label{eq:tilda_mu}
\tilde{\mu}_q = \frac{1}{V_\mathrm{c}}\int{\rm d}^3\pmb{r}\, \frac{\delta E_{\rm ETF}}{\delta n_q(\pmb{r})} \, . 
\eeqy 
This shows that $\tilde{\mu}_q$ does not generally coincide with $\mu_q$ to the extent that the Euler-Lagrange Eqs.~(\ref{eq:Euler-Lagrange}) are not exactly fulfilled
except in the limit of homogeneous matter. Alternative expressions for the chemical potentials can be obtained from Eqs.~\eqref{eq:delta_E_ETF_param2} and \eqref{eq:delta_E_ETF_param4}: 
\beqy\label{eq:chemical-potential-restricted1}
\tilde{\mu}_q = \dfrac{\displaystyle\int{\rm d}^3\pmb{r}\, \dfrac{\partial f_q(r,\pmb{x_q},R)}{\partial \pmb{x_{q}}} \dfrac{\delta E_{\rm ETF}}{\delta n_q(\pmb{r})} } {\displaystyle \int{\rm d}^3\pmb{r}\, \dfrac{\partial f_q(r,\pmb{x_q},R)}{\partial \pmb{x_{q}}} }
\, ,
\eeqy 
\beqy\label{eq:chemical-potential-restricted3}
\tilde{\mu}_q = \dfrac{\displaystyle\int{\rm d}^3\pmb{r}\,f_q(r,\pmb{x_q},R) \dfrac{\delta E_{\rm ETF}}{\delta n_q(\pmb{r})} } {\displaystyle \int{\rm d}^3\pmb{r}\, f_q(r,\pmb{x_q},R) }
\, .
\eeqy 
These expressions can be used to check the numerical accuracy of the code and the convergence of the solution. 
Note that $R$ is not treated here as a free parameter since it is understood that the volume of the cell is fixed.

To a very good approximation, electrons can be assumed to be uniformly distributed (see, e.g., Ref.~\cite{Maruyama2005}). This amounts to 
parametrizing the electron density using a profile similar to Eq.~\eqref{eq:parametrized-profiles1} with a pure background term, i.e. 
$n_{Be}=n_e$ and $n_{\Lambda e}=0$. It thus follows from Eq.~\eqref{eq:tilda_mu} that the electron chemical potential is given by
\beqy\label{eq:electron-chemical-potential-restricted}
\tilde{\mu}_e = \frac{d \mathcal{E}_e}{d n_e} + \frac{e^2}{2} \left(\frac{3}{\pi}\right)^{1/3}n_e^{1/3}-  \frac{e}{V_{\rm c}} \int{\rm d}^3\pmb{r}\, U_{\rm Coul}(\pmb{r}) \, .
\eeqy
The integral can be calculated and expressed in terms of the mean square radius of the proton distribution defined by
\beqy\label{eq:mean-square-proton-radius}
\langle r^2\rangle = \frac{4\pi}{N^{(\mathrm{c})}_e}\int_0^R {\rm d}r\, n_p(r) r^4 \, .
\eeqy
The electron chemical potential is finally given by 
\beqy\label{eq:electron-chemical-potential-restricted-final}
\tilde{\mu}_e = \frac{d \mathcal{E}_e}{d n_e} + \frac{e^2}{2} \left(\frac{3}{\pi}\right)^{1/3}n_e^{1/3}- \frac{3}{10} \frac{N_p^{\mathrm{(c)}} e^2}{R} \left(1-\frac{5}{3}\frac{\langle r^2\rangle}{R^2}\right) \, .
\eeqy

Instead of introducing Lagrange multipliers and minimizing the grand potential, one can alternatively fix the numbers of particles and 
minimize the energy. This leads to the same expressions for the chemical potentials, as shown in Appendix~\ref{app:chemical-potential}.

\section{Pressure formulas} 
\label{sect:pressure}

\subsection{General expression}

As shown in Section~\ref{sec:general-considerations}, the pressure can be calculated from the energy 
of a Wigner-Seitz cell. In general, the energy is a functional of the 
nucleon single-particle wave functions
in the HF theory or of the local densities in the ETF approximation. However, 
it is understood here that the pressure is calculated for the \emph{optimum} single-particle 
wave functions or local densities, i.e. those minimizing the HF or ETF energy $E_\mathrm{c}$ for 
fixed $N^{(\mathrm{c})}_i$

In the approximation of spherical Wigner-Seitz cells, Eq.~\eqref{eq:pressure-cell} reads 
\beqy
P =-\frac{1}{4\pi R^2}\left(\frac{\partial E_\mathrm{c}}{\partial R}\right)_{N^{(\mathrm{c})}_q} 
\, .
\eeqy
The total numbers of nucleons and electrons in the cell are given by 
\bmlet
\beqy\label{B4a}
N^{(\mathrm{c})}_q=4\pi \int_0^R {\rm d}r\, r^2 n_q(r)\, ,
\eeqy 
\beqy\label{B4b}
N^{(\mathrm{c})}_e=4\pi \int_0^R {\rm d}r\, r^2 n_e(r)\, .
\eeqy 
\emlet
The energy in the cell can be explicitly written as 
\beqy
E_\mathrm{c}=4\pi \int_0^R {\rm d} r\, r^2 {\mathcal E}(r)\, .
\eeqy

Differentiating $E_\mathrm{c}$ with respect to $R$ thus yields 
\beqy\label{eq:pressure-variations}
P=-\mathcal{E}(R)- \frac{1}{R^2}\int_0^R{\rm d}r\, r^2 
\sum_q\biggl[ \frac{\delta E}{\delta n_q(r)}\frac{\partial n_q(r)}{\partial R}
+\frac{\delta E}{\delta \tau_q(r)}\frac{\partial \tau_q(r)}{\partial R}
+\frac{\delta E}{\delta J_q(r)}\frac{\partial J_q(r)}{\partial R}\biggr]\notag \\ 
- \frac{1}{R^2}\int_0^R{\rm d}r\, r^2 \frac{\delta E}{\delta n_e(r)}\frac{\partial n_e(r)}{\partial R}\, .
\eeqy

Let us recall that the pressure must be calculated keeping the number of each type of 
particles fixed. Using Eqs.~(\ref{B4a}) and (\ref{B4b}), the following identities must 
be fulfilled: 
\bmlet
\beqy\label{B7a}
\int_0^R {\rm d} r\, r^2 \frac{\partial n_q(r)}{\partial R}=-R^2 n_q(R)\, ,
\eeqy
and
\beqy\label{B7b}
\int_0^R {\rm d} r\, r^2 \frac{\partial n_e(r)}{\partial R}=-R^2 n_e(R)\, . 
\eeqy
\emlet

Note that Poisson's equation can be analytically solved in the spherical Wigner-Seitz cell approximation and the Coulomb potential reads 
\beqy \label{eq:Coulomb-potential}
U_{\rm Coul}(r)=2\pi e \int_0^R dr' r'^2 n_{\rm ch}(r')\frac{r+r'-|r-r'|}{rr'} 
\eeqy 
(see, e.g., Ref.~\cite{Pearson+12}). 
The variation of the direct Coulomb energy with respect to $R$ then formally yields a pressure term $-2\mathcal{E}_{\rm Coul,dir}(R)$; the factor of $2$ arises from the explicit dependence of the Coulomb potential $U_{\rm Coul}(r)$ on $R$. However, $\mathcal{E}_{\rm Coul,dir}(R)=0$ from the electric charge neutrality of the Wigner-Seitz cell.

\subsection{Hartree-Fock theory}

As shown in Appendix~\ref{app:pressure-HF}, the pressure in the HF theory can be expressed as 
\beqy\label{eq:pressure-HF-approx}
P_{\rm HF}=-\mathcal{E}_{\rm HF}(R)+
\frac{1}{4\pi }\sum_q\sum_{\alpha(q)}g^{(q)}_\alpha \epsilon^{(q)}_\alpha \mathcal{R}^{(q)}_{\alpha}(R)^2 +\tilde{\mu}_e n_e \, .
\eeqy
Here, we have neglected electron charge screening effects by considering $n_e(r)\approx n_e$. 
Using the energy decomposition~\eqref{eq:energy-contributions}, the pressure can be equivalently written as 
\beqy\label{eq:pressure-HF-approx2}
P_{\rm HF}=P_e+P_\mathrm{Coul,dir}+P_\mathrm{Coul,exch} +P_{\rm nuc} \, ,
\eeqy
where $P_e$ is the pressure of an ideal relativistic electron Fermi gas of density $n_e$, 
\beqy 
P_\mathrm{Coul,dir}\equiv -\frac{3 e n_e}{R^3}\int_0^R {\rm d}r\, r^2 U_\mathrm{Coul}(r)
\eeqy 
and 
\beqy
P_{\rm Coul, exch} \equiv  \frac{e^2}{8} \left(\frac{3}{\pi}\right)^{1/3}n_e^{4/3}-x \frac{e^2}{4} \left(\frac{3}{\pi}\right)^{1/3}n_{p}(R)^{4/3} 
\eeqy
account for the direct and exchange Coulomb terms respectively, 
while the purely nuclear contribution is given by 
\beqy 
P_\mathrm{nuc}\equiv 
\frac{1}{4\pi }\sum_q\sum_{\alpha(q)}g^{(q)}_\alpha \epsilon^{(q)}_\alpha \mathcal{R}^{(q)}_{\alpha}(R)^2  -\mathcal{E}_{\rm nuc}(R)\, .
\eeqy 
We have made use of the fact that $U_{\rm Coul}(R)=0$ as a consequence of global electric charge neutrality. 

In the outer crust of neutron stars, where all nucleons are tightly bound inside 
nuclei, the nucleon wavefunctions are exponentially suppressed outside therefore 
$\mathcal{R}^{(q)}_{\alpha}(R)\approx 0$. 
The pressure is then essentially determined by electrons with electrostatic corrections, 
as expected. In the inner crust, protons remain bound inside clusters 
(except possibly in the very deep layers consisting of nuclear pastas), but
some neutrons are free and therefore yield a finite contribution to the pressure. 

In practice, Eq.~\eqref{eq:pressure-HF-approx2} is very sensitive to the choice of 
the boundary conditions, and therefore may not be very accurate. In particular, 
with the Dirichlet boundary conditions~\eqref{eq:Dirichlet}, the purely nuclear 
contribution to the pressure vanishes. 
For a given composition, the pressure of dense inhomogeneous matter 
with free neutrons is thus strongly underestimated even 
if the total energy is hardly 
altered. This spurious suppression of the pressure will change the equilibrium composition 
of the inner crust, therefore also transport properties such as the electric and thermal 
conductivities, as well as elastic constants. With the Neumann boundary condition~\eqref{eq:Neumann}, 
the nuclear contribution to the pressure remains finite but is still prone to systematic 
errors due to the spurious discretization of unbound single-particle energies $\epsilon^{(q)}_\alpha$. 
Because this discretization also directly impacts the neutron chemical potential (and that of protons 
beyond proton drip) as well as the energy $E_\mathrm{HF}$ of the cell, the alternative pressure 
formula~\eqref{eq:pressure-general2} is subject to similar errors. 

This analysis shows that the errors incurred by the choice of approximate 
boundary conditions can propagate to global thermodynamic properties. Without sacrificing 
the Wigner-Seitz approximation, the equation of state can be more reliably calculated within the ETF 
approach. 

\subsection{Extended Thomas Fermi approach: unrestricted minimization}

In the ETF approximation, the energy depends neither on $\tau_q(\pmb{r})$ nor 
$\pmb{J_q}(\pmb{r})
$ therefore the general pressure formula~\eqref{eq:pressure-variations} reduces to 
\beqy\label{eq:pressure-variations-ETF}
P_{\rm ETF}=-\mathcal{E}_{\rm ETF}(R)- \frac{1}{R^2}\int_0^R{\rm d}r\, r^2 
\biggl[\sum_q \frac{\delta E_{\rm ETF}}{\delta n_q(r)}\frac{\partial n_q(r)}{\partial R}
+\frac{\delta E_{\rm ETF}}{\delta n_e(r)}\frac{\partial n_e(r)}{\partial R}\biggr]\, .
\eeqy
Let us recall that the pressure must be calculated for the \emph{optimum} nucleon 
density profiles.  

If the ETF energy is minimized without any restriction on the local densities, 
it can be easily seen from the Euler-Lagrange equations~\eqref{eq:Euler-Lagrange} and \eqref{eq:Euler-Lagrange-electrons} using Eqs.~\eqref{B7a} and \eqref{B7b} that the 
pressure can be expressed in this case as 
\beqy\label{eq-pressure-Euler-Lagrange}
P_{\rm ETF}=-\mathcal{E}_{\rm ETF}(R)+\mu_e n_e(R)+\sum_q \mu_q n_q(R)\, .
\eeqy
This simple formula was derived in Ref.~\cite{Pearson+12} and can be considered as
a generalization of the expression derived in atomic 
physics in the framework of the Thomas-Fermi-Dirac model (see, e.g., 
 Ref.~\cite{AbrahamsShapiro1990} and references therein). Although this expression resembles 
to that of a homogeneous background mixture of nucleons and electrons with 
densities $n_q(R)$ and $n_e(R)$ respectively, the effects of the clusters are 
taken into account but are hidden in the chemical potentials and in the 
higher-order derivatives of the nucleon distributions entering the energy density (only the 
first derivatives are required to vanish at the border of the cell). 

With the pressure formula~\eqref{eq-pressure-Euler-Lagrange}, we can rewrite the effective chemical potential of clusters~\eqref{eq:chemical-potential-cluster-cell-explicit} as
\beqy \label{eq:chemical-potential-cluster-decomposition}
\mu_N=\mu_X - \mu_e N^X_e - \sum_q \mu_q N^X_q\, ,
\eeqy 
where $N^X_i\equiv N_i^{\mathrm{c}}-V_\mathrm{c} n_i(R)$ is defined as the number of particles in each cluster, and 
\beqy\label{eq:nuclei-chemical-potential-def} 
\mu_X \equiv E_{\rm ETF} - V_\mathrm{c} \, \mathcal{E}_{\rm ETF}(R) 
\eeqy 
their `true' chemical potential.

In the outer crust, where protons and neutrons are bound inside nuclei we have $N_p^X=N_p\equiv Z$ and $N_n^X=N_n\equiv A-Z$. Ignoring electron charge screening effects $N_e^X=0$ and Eq.~\eqref{eq:chemical-potential-cluster-decomposition} thus reduces to
\beqy\label{eq:cluster-chemical-potential-outer-crust}
\mu_N=\mu_X - Z \mu_p - (A-Z)\mu_n \, . 
\eeqy 
As shown in Appendix~\ref{app:chemical-potential-outer-crust}, the chemical potential~\eqref{eq:nuclei-chemical-potential-def} can be written as    
\beqy\label{eq:chemical-potential-nuclei} 
\mu_X = M(A,Z)c^2 - \frac{9}{10} \frac{Z^2 e^2}{R} \, ,
\eeqy 
where $M(A,Z)$ is the nuclear mass in vacuum (including nuclear energy as well as direct and exchange proton-proton Coulomb energies), $c$ is the speed of light and the second term is a correction due to electron-electron and electron-proton interactions in the stellar medium neglecting the finite size of the nuclei. 
This shows that the effective chemical potential $\mu_N$ actually represents the chemical affinity of all possible  
nuclear reactions with separate conservations of neutrons and protons and
effectively summarized by 
\beqy\label{eq:nuclear-reactions}
^A_ZX \leftrightarrow Z p + (A-Z) n \, .
\eeqy 
The condition $\mu_N=0$ therefore corresponds to the equilibrium with respect to strong but not necessarily 
weak reactions. The latter (beta) equilibrium condition is embedded in Eq.~\eqref{eq:beta-equilibrium}. 
In accreting neutron stars, nuclear reactions~\eqref{eq:nuclear-reactions} are frozen by the Coulomb barrier  (except possibly for 
pycnonuclear fusions for the lightest elements; 
see, e.g., Ref.~\cite{Yakovlev_ea06}), but clusters could still 
capture/emit neutrons and electrons.
Therefore, only the beta-equilibrium condition~\eqref{eq:beta-equilibrium} applies (see Refs.~\cite{GC20_DiffEq,GKC_psi21,SGC22a,SGC23_compos,GC24} for details).
In nonaccreted neutron stars, matter is generally assumed to be cold and fully catalyzed, i.e. in its absolute ground state at zero temperature. Therefore, both strong and weak equilibrium conditions are satisfied. 
For the sake of clarity, let us stress that the equilibrium~\eqref{eq:nuclear-reactions} does not mean that the underlying contributing reactions really occur (as for accreted crusts, strong reactions are blocked by Coulomb barrier).  Rather, it means that the reaction~\eqref{eq:nuclear-reactions} has zero energy gain because the equilibrium was reached at a preceding stage of neutron star evolution, during which the temperature was still high enough to overcome the Coulomb barrier.

\subsection{Extended Thomas Fermi approach: parametrized profiles}
\label{sect:pressure:restricted-ETF}

The pressure formula~\eqref{eq-pressure-Euler-Lagrange} was demonstrated in Appendix B of Ref.~\cite{Pearson+12} within the ETF approach and applied  there
using 
the parametrized profiles 
of Ref.~\cite{Onsi+08} with
\beqy\label{eq:parametrized-profiles3}
f^\mathrm{strD}_q(r) = \frac{1}{1 +\exp \left[\Big(\dfrac{C_q - R}
{r - R}\Big)^2 - 1\right]\exp \Big(\dfrac{r-C_q}{a_q}\Big) } \quad ,
\eeqy
and considering uniformly distributed electrons of density $n_e$. 
It was remarked, 
that all the derivatives of the nucleon densities $n_q(r)$ vanish at the cell edge $r=R$, 
thus the nucleon chemical potentials~\eqref{eq:Euler-Lagrange} evaluated at $r=R$ 
can be reduced to those of homogeneous nuclear matter with neutron density $n_n(R)=n_{Bn}$ and proton density $n_p(R)=n_{Bp}$. 
As a result, the final expression for pressure in Ref.~\cite{Pearson+12}
was written as a sum of 'homogeneous' matter partial pressures
\beqy\label{eq-pressure-Euler-Lagrange-app}
P_{\rm hom}= P_e+ P_{\rm nuc}+ P_{\rm Coul,exch}\, ,
\eeqy
where $P_e$ is the pressure of 
an ideal relativistic electron Fermi gas of density $n_e(R)=n_e$, 
$P_{\rm nuc}$ is the pressure of homogeneous nuclear matter with neutron density $n_n(R)=n_{Bn}$ 
and proton density $n_p(R)=n_{Bp}$, and 
\beqy
P_{\rm Coul, exch} =  \frac{e^2}{8} \left(\frac{3}{\pi}\right)^{1/3}n_e(R)^{4/3}-x \frac{e^2}{4} \left(\frac{3}{\pi}\right)^{1/3}n_{p}(R)^{4/3} 
\eeqy
accounts for the Coulomb exchange. 
The formula~\eqref{eq-pressure-Euler-Lagrange-app} was applied in our previous studies, and in 
particular in our calculations of unified equations of state~\cite{Pearson_ea18_bsk22-26}. 
However, Eq.~\eqref{eq-pressure-Euler-Lagrange} does not strictly hold if the densities are 
parametrized because the Euler-Lagrange equations~\eqref{eq:Euler-Lagrange} are not exactly fulfilled in this 
case. As a matter of fact, it can be noticed that Eq.~\eqref{eq-pressure-Euler-Lagrange-app} 
does not contain any contribution from direct Coulomb interactions. In particular, the well-known 
lattice pressure (see, e.g., Ref.~\cite{ShapiroTeukolsky}) is missing. 

The correct pressure formula is derived in Appendix~\ref{app:pressure-ETF-param}, and can be written as
\beqy\label{eq-pressure-Euler-Lagrange-corrected-general}
\tilde{P}_{\rm ETF}=P_e+P_\mathrm{nuc}+P_{\rm Coul,dir}+ P_{\rm Coul,exch}+\delta \tilde{P}_\mathrm{nuc} + \delta P^\nabla_\mathrm{nuc} \, ,
\eeqy
where $P_{\rm Coul,dir}$ is the lattice pressure given by 
\beqy \label{eq:lattice-pressure}
P_{\rm Coul,dir}=-\frac{2 \pi}{5} e^2 n_e^2 R^2 \left(1-\frac{5}{3}\frac{\langle r^2\rangle}{R^2}\right)\, ,
\eeqy 
\beqy\label{eq-pressure-Euler-Lagrange-corrected-param}
\delta \tilde{P}_\mathrm{nuc} \equiv \sum_q n_{q}(R) \biggl[ \tilde{\mu}_q -\frac{\delta E_{\rm ETF}}{\delta n_q(r)}\biggr\vert_{r=R}\biggr]  + \frac{1}{R^2}\sum_q n_{\Lambda q} \int_0^R{\rm d}r\, r^2 \ \biggl[\tilde{\mu}_q-\frac{\delta E_{\rm ETF}}{\delta n_q(r)}\biggr]\frac{\partial f_q(r)}{\partial R}\biggr\vert_{\pmb{x_q}} 
\eeqy
is the correction due to the parametrization of the nucleon distributions, while 
\beqy \label{eq-pressure-Euler-Lagrange-corrected-derivatives}
\delta P^\nabla_\mathrm{nuc}\equiv -\mathcal{E}^\nabla_\mathrm{nuc}(R)+\sum_q n_q(R)\biggl[\frac{\delta E_{\rm ETF}}{\delta n_q(r)}\biggr\vert_{r=R}-\mu^\mathrm{hom}_q\biggr] 
\eeqy 
is the correction due to density derivatives, whose contribution to the nuclear energy density $\mathcal{E}_\mathrm{nuc}(r)$ is denoted by $\mathcal{E}^\nabla_\mathrm{nuc}(r)$ (by definition this term vanishes 
in the homogeneous nuclear matter). Here $\delta E_\mathrm{ETF}/\delta n_q(r)$ is evaluated for the parametrized profile, and $\mu^\mathrm{hom}_q$ is the nucleon chemical potential given by Eq.~\eqref{eq:muq_HF} in homogeneous nuclear matter with neutron density $n_{n}(R)$ and proton density $n_{p}(R)$.

The pressure formula~\eqref{eq-pressure-Euler-Lagrange-corrected-general} is amenable to further simplifications for parametrized profiles such as ~\eqref{eq:parametrized-profiles3} for which the first four derivatives of $n_q(\pmb{r})$ all vanish at the cell edge\footnote{For the profile parametrization~\eqref{eq:parametrized-profiles3}, all derivatives of $n_q(\pmb{r})$ actually vanish at the cell edge. As we have recently shown~\cite{ShchechilinParam+23}, this requirement is too strong in the innermost layers of the crust.}. 
First, $\mathcal{E}^\nabla_\mathrm{nuc}(R)=0$ and $\delta E_\mathrm{ETF}/\delta n_q(r)\vert_{r=R}$ reduces to $\mu^\mathrm{hom}_q$ therefore $\delta P^\nabla_\mathrm{nuc}=0$ and the pressure formula~\eqref{eq-pressure-Euler-Lagrange-app} can be written in the simplified form
\beqy\label{eq-pressure-Euler-Lagrange-corrected-str-damping}
\tilde{P}_{\rm ETF}=P_\mathrm{hom} + P_\mathrm{Coul,dir}+ \delta \tilde{P}_\mathrm{nuc} \, ,
\eeqy 
where 
\beqy\label{eq-pressure-Euler-Lagrange-corrected-param-str-damping}
\delta \tilde{P}_\mathrm{nuc} = \sum_q n_{q}(R) (\tilde{\mu}_q -\mu^\mathrm{hom}_q)  + \frac{1}{R^2}\sum_q n_{\Lambda q} \int_0^R{\rm d}r\, r^2 \ \biggl[\tilde{\mu}_q-\frac{\delta E_{\rm ETF}}{\delta n_q(r)}\biggr]\frac{\partial f_q(r)}{\partial R}\biggr\vert_{\pmb{x_q}} \, .
\eeqy
Secondly, the vanishing of the derivatives facilitates the calculation of the nucleon chemical potential $\tilde{\mu}_q$, which enters into the expression of $\delta \tilde{P}_\mathrm{nuc}$. Namely,  Eq.~\eqref{eq:tilda_mu} after substitution of Eq.~\eqref{eq:Euler-Lagrange} and integrations by parts leads to
\beqy\label{eq:tilda_mu_str_damping} 
\tilde{\mu}_q=\frac{3}{R^3} \int_0^R {\rm d}r\, r^2  \frac{\partial \mathcal{E}_{\rm ETF}(r)}{\partial n_q(r)}  \, .
\eeqy 
Note that the derivatives are only supposed to vanish at the border of the cell. Therefore, $\partial \mathcal{E}_{\rm ETF}(r)/\partial n_q(r)$ not only contains the leading TF expression but also higher order corrections. 

\subsection{Numerical comparisons}

To assess the 
accuracy of the various formulas derived above, we have performed numerical calculations in a neutron-star crust for different average baryon densities ranging
from $0.001$~fm$^{-3}$ to $0.08$~fm$^{-3}$ using the same functional BSk24~\cite{Goriely_ea_Bsk22-26} and the same profile parametrization~\eqref{eq:parametrized-profiles1} with \eqref{eq:parametrized-profiles3}, as in our previous calculations of unified equations of state of nonaccreted neutron stars~\cite{Pearson_ea18_bsk22-26,Pearson+22}. For simplicity, we have truncated the ETF expansion to the leading order. Within this TF approach, we have obtained the optimal nucleon distributions inside the optimal spherical Wigner-Seitz cells. The proton density at the border of the cell is found to remain negligibly small for densities $\bar{n}\lesssim0.077$\,fm$^{-3}$ (the crust-core transition occurs at $\bar{n}_\mathrm{cc}\sim0.08$\,fm$^{-3}$). In other words, $n_p(\pmb{r})$ could have well been determined using a 
profile with $n_{Bp}=0$. Therefore, the derivative $\partial n_p(\pmb{r})/\partial n_{Bp}$  is ill-defined
and the expression~\eqref{eq:tilda_mu} derived from Eq.~\eqref{eq:delta_E_ETF_param1} is no longer valid. 
We find an excellent agreement among all the remaining 
expressions of the proton chemical potentials given by Eqs.~\eqref{eq:chemical-potential-restricted1}-\eqref{eq:chemical-potential-restricted3} within the precision of the code. For neutrons,  
Eqs.~\eqref{eq:tilda_mu}-\eqref{eq:chemical-potential-restricted3} are all found to be numerically equivalent. 
The values we obtain also coincide with the basic definition~\eqref{eq:muq_def} evaluated numerically at a higher computational cost using finite differences on specially calculated dense grid (additional calculations were needed to adjust the step size and ensure the reliability of the results for each density $\bar n$). These nucleon chemical potentials have allowed us to check that the equilibrium condition~\eqref{eq:beta-equilibrium} indeed holds. With these comparisons, we have further tested the numerical accuracy of the minimization procedure of our ETF code used in Ref.~\cite{Shchechilin+sym23}. Using the nucleon chemical potentials thus obtained, we have computed the pressure from Eq.~\eqref{eq:pressure-catalyzed}. 
It is found to be in perfect agreement with the pressure~\eqref{eq:pressure-general} evaluated numerically using finite differences  calculated with an optimized grid. 
Moreover, the corrected pressure formula~\eqref{eq-pressure-Euler-Lagrange-corrected-str-damping} yields exactly the same values within numerical precision. Those formulas can be easily implemented in existing computer codes thus allowing for more accurate and computationally much faster evaluations of the chemical potentials and pressure without the need to calibrate grids at each density $\bar n$. 

\begin{figure}
	\includegraphics[width=0.5\columnwidth]{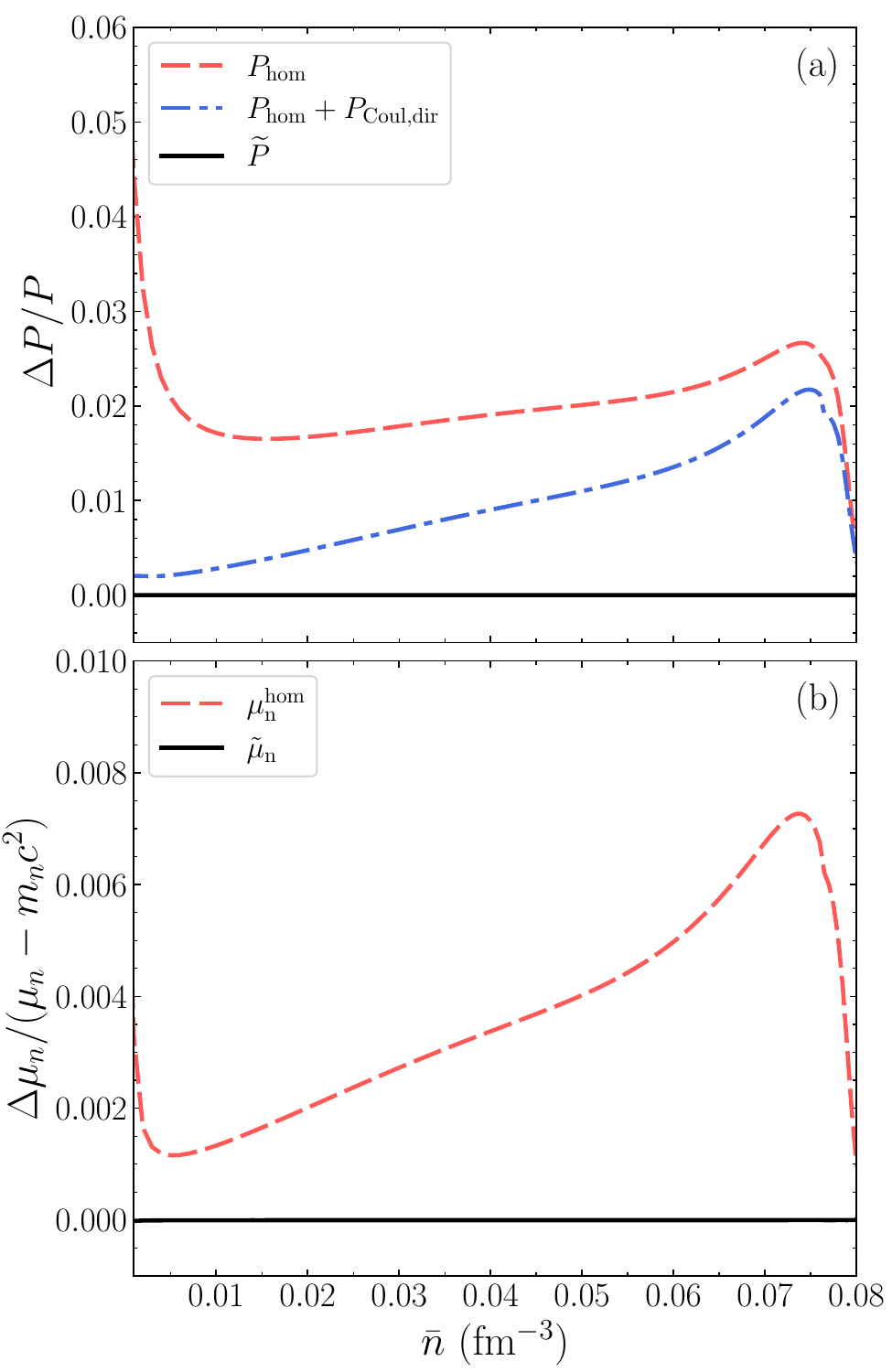}
	\caption{Panel (a): relative deviations between the following formulas for the pressure and the numerical evaluation of Eq.~\eqref{eq:pressure-general2} in the inner crust of a nonaccreted neutron star as a function of the average baryon number density $\bar n$ in fm$^{-3}$: Eq.~\eqref{eq:pressure-catalyzed} with $\tilde{\mu}_\mathrm{n}$ from Eq.~\eqref{eq:tilda_mu_str_damping} (black solid line), Eq.~\eqref{eq-pressure-Euler-Lagrange-app}  with (blue dash-dotted line) and without (red dashed line) the lattice correction~\eqref{eq:lattice-pressure}. Panel (b): 
    relative deviations between the neutron chemical potential (with the bare mass subtracted out) 
    obtained from Eq.~\eqref{eq:tilda_mu_str_damping} (black solid line) or from Eq.~\eqref{eq:muq_HF} with neutron density $n_{n}(R)$ and proton density $n_{p}(R)$ (red dashed line) and that calculated numerically from Eq.~\eqref{eq:muq_def}. 
    The composition and nucleon distributions were calculated in Ref.~\cite{Shchechilin+sym23} within the 4th-order ETFSI method using the profile parametrization given by Eq.~\eqref{eq:parametrized-profiles3}. Small jumps in the slope of the curves around $\bar{n}\approx0.076$~fm$^{-3}$ correspond to the proton-drip point. See text for details.}
	\label{fig:Pmun}
\end{figure}

Next, we have used our previous full 4th-order ETF results for the inner crust composition and nucleon distributions obtained with BSk24 including SI corrections~\cite{Shchechilin+sym23}.  
As shown in Sec.~\ref{sect:pressure:restricted-ETF}, for the parametrization~\eqref{eq:parametrized-profiles3} the neutron chemical potential $\tilde{\mu}_n$ can be relatively easily calculated from Eq.~\eqref{eq:tilda_mu_str_damping}. 
We have found it to be in perfect agreement with the neutron chemical potential~\eqref{eq:muq_def} obtained numerically from the ETF energy using finite differences, as can be seen  
in Fig.~\ref{fig:Pmun}. With this $\tilde{\mu}_n$, we have calculated the pressure
$\tilde{P}$ from Eq.~\eqref{eq:pressure-catalyzed} using the ETF energy density (as discussed in Sec.~\ref{sec:general-considerations}, this formula is still applicable even when $N_p^{\mathrm{(c)}}$ is given by the SI corrections). 
Again, the results coincide  with 
those obtained from the appropriate numerical differentiation of the ETF energy within 
numerical accuracy, as displayed in Fig.~\ref{fig:Pmun} (the slight change of slope around $\bar{n}\approx0.076$\,fm$^{-3}$ is caused by the proton drip point). For comparison, we have also computed the pressure using the formula~\eqref{eq-pressure-Euler-Lagrange-app} applied in our previous studies. This pressure is found to be systematically too high (by a few percents), as can be seen in the same figure. At very low densities $\bar n$, the discrepancy mainly comes from the omission of the (negative) lattice pressure~\eqref{eq:lattice-pressure}. 
At higher densities, the additional correction $\delta \tilde{P}_\mathrm{nuc}$
in Eq.~\eqref{eq-pressure-Euler-Lagrange-corrected-str-damping} becomes more and more  
important. This stems from the fact that the background nucleon densities and the
volume fraction filled by clusters both increase.
Figure~\ref{fig:Pmun} also indicates that the neutron chemical potential calculated at the cell border\footnote{This equality stems from the fact that all derivatives vanish at the border for the parametrization adopted in Ref.~\cite{Shchechilin+sym23}}, $\mu_\mathrm{n}(R)=\mu_\mathrm{n,hom}$, differs from the neutron chemical potential calculated from numerical derivative by $1\%$ at most throughout the considered region.

The errors on the pressure propagate to the global properties of the neutron star. To estimate these errors, we have solved the Tolman-Oppenheimer-Volkoff equations~\cite{Tolman39,OV39} using our previous pressure formula~\eqref{eq-pressure-Euler-Lagrange-app} and comparing with the new one~\eqref{eq-pressure-Euler-Lagrange-corrected-str-damping}. 
The resulting increase in the neutron-star radius of the order of ten meters for a $1.4 M_\odot$
is quite marginal and well below the current observational uncertainties. However, such an accuracy could be reached with next generation gravitational-wave detectors~\cite{Huxford2024}. 

\begin{figure}
	\includegraphics[width=0.5\columnwidth]{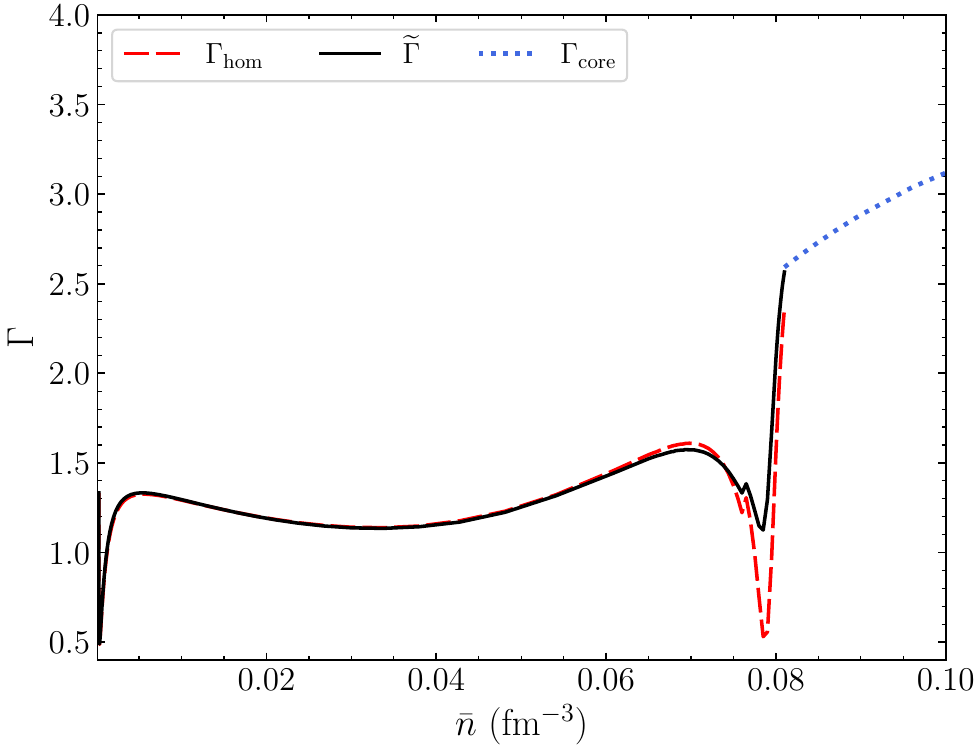}
	\caption{Adiabatic index in the inner crust 
 (black and red curves) and core (blue curve) of a nonaccreted neutron star as a function of the average baryon number density $\bar n$ in fm$^{-3}$. The black curve was obtained from Eq.~\eqref{eq:pressure-catalyzed} with $\tilde{\mu}_\mathrm{n}$ from Eq.~\eqref{eq:tilda_mu_str_damping}. The red curve was obtained from Eq.~\eqref{eq-pressure-Euler-Lagrange-app}. The blue dotted curve was calculated for $\beta$-equilibrated homogeneous $npe$ matter.}
	\label{fig:Gamma}
\end{figure}

\begin{figure}
	\includegraphics[width=0.5\columnwidth]{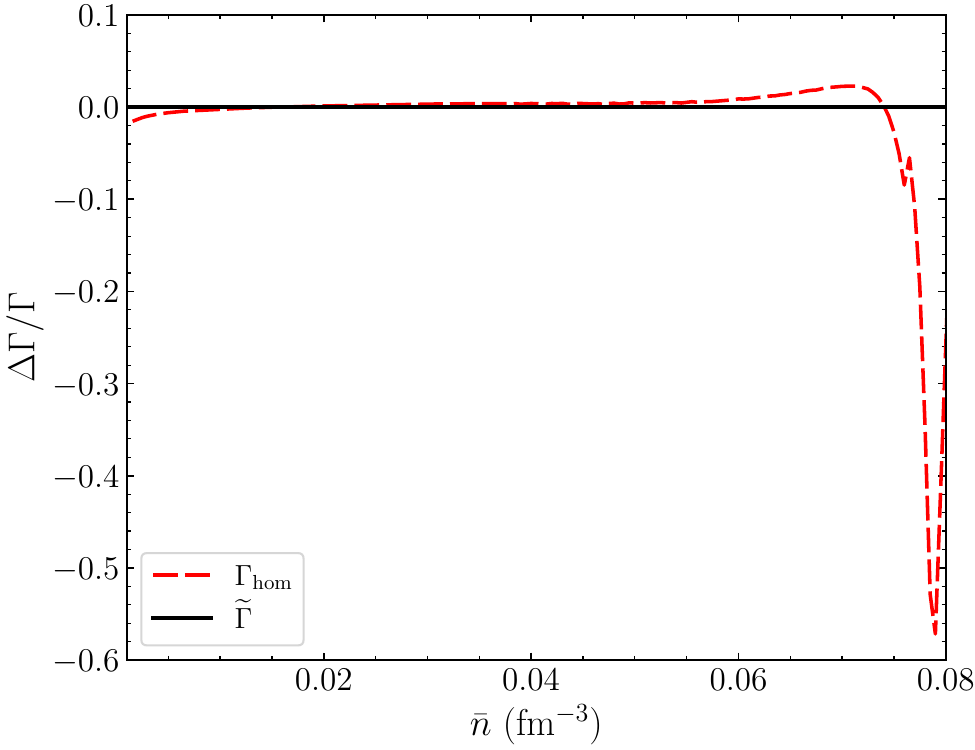}
	\caption{Relative deviation in the adiabatic index in the inner crust of a nonaccreted neutron star as a function of the average baryon number density $\bar n$ in fm$^{-3}$. The legend is the same as in Fig.~\ref{fig:Gamma}.}
	\label{fig:Gamma-error}
\end{figure}

The errors are more significant for the adiabatic index defined by 
\beqy\label{eq:adiabatic-index-def}
\Gamma = \frac{\bar{n}}{P}\frac{dP}{d\bar{n}} \, ,
\eeqy 
As shown in Fig.~\ref{fig:Gamma},
the use of the approximate pressure formula~\eqref{eq-pressure-Euler-Lagrange-app} leads to a much more pronounced drop of the adiabatic 
index around $\overline n\lesssim 0.08$~fm$^{-3}$: $\Gamma$ is about a factor of 2 smaller. This artificial drop is comparable to the physical 
drop associated with the neutron-drip transition delimiting the outer and inner crusts. The approximate formula thus mimics a spurious 
interface within the inner crust. The relative errors throughout the inner crust are displayed in Fig.~\ref{fig:Gamma-error}. They may have 
a non-negligible impact on the oscillation spectrum, but we left respective calculations for future studies.

\section{Conclusions}

We have obtained very general pressure formulas applicable to the outer and inner crusts of a cold 
nonaccreted or accreted neutron star within the spherical Wigner-Seitz cell approximation 
both in the HF and ETF methods using  effective interactions of the Skyrme type, respectively in 
Eqs.~\eqref{eq:pressure-HF-approx} and \eqref{eq-pressure-Euler-Lagrange}. Quite remarkably, these 
formulas depend on the values of the nucleon single-particle wavefunctions or of the nucleon densities 
at the border of the Wigner-Seitz cell. This shows that the boundary conditions directly impact the 
calculation of the equation of state, and should thus be very carefully chosen. Indeed, requiring 
the vanishing of the nucleon wavefunctions within the HF approach as in 
Refs.~\cite{Pizzochero+2002,Barranco+2010,Baroni+2010}, leads to the exact cancellation of the nucleon 
contribution to the pressure. Imposing the vanishing of the first derivative of the wavefunctions or 
mixed boundary conditions depending on parity avoids this nonphysical cancellation of the nucleon pressure. 
But spurious contributions are still present and can hardly be removed because of the nonlinearity of 
the HF equations. The errors propagate throughout the self-consistent calculations and can alter the 
equilibrium composition of the crust therefore also transport and elastic properties. This problem does 
not arise in the ETF approach since it deals only with local nucleon densities and only the first derivatives 
of the nucleon densities are generally required to vanish. On the other hand, our analysis has demonstrated 
that the TF method must be extended to properly describe bound nucleons. 

We have also calculated the pressure in the restricted ETF approach with parametrized nucleon density 
distributions; the result is given by Eq.~\eqref{eq-pressure-Euler-Lagrange-corrected-general}. This formula 
corrects the one we derived in Ref.~\cite{Pearson+12}: it now includes i) the lattice pressure~\eqref{eq:lattice-pressure} thus improving the description of the transition between the outer and inner crusts, ii) the 
correction~\eqref{eq-pressure-Euler-Lagrange-corrected-param} due to the restrictions imposed by the 
parametrizations, and iii) the correction~\eqref{eq-pressure-Euler-Lagrange-corrected-derivatives} due to 
the non-vanishing of higher-order derivatives of the nucleon densities at the border of the cell. 
For cold dense matter in full equilibrium, we have shown that the pressure reduces to a very simple 
formula~\eqref{eq:pressure-catalyzed}. 

The nucleon chemical potentials also require special treatment in the restricted ETF approach. It is 
because the parametrized nucleon density distributions generally do not satisfy the Euler-Lagrange equations~\eqref{eq:Euler-Lagrange}. Thus, if these nucleon density distributions are applied to determine 
the nucleon chemical potentials, the results may be unreliable, being dependent on the chosen position inside 
the cell.
We have shown how to obtain thermodynamically consistent chemical potentials and we have derived equivalent 
expressions, as many as the number of parameters used to describe the nucleon density distributions. We have 
demonstrated explicit formulas, namely Eqs.~\eqref{eq:delta_E_ETF_param1}-\eqref{eq:delta_E_ETF_param4}, for 
generic profiles, including those introduced in Refs.~\cite{Onsi+08,ShchechilinParam+23}. These equivalent 
expressions can also be implemented to test the convergence of the solutions and to assess the numerical accuracy 
of the code. Moreover, the realism of the adopted parametrization can be more rigorously measured by comparing 
these chemical potentials with those obtained from the Euler-Lagrange equations~\eqref{eq:Euler-Lagrange}.

Using those formulas and the inner-crust composition previously obtained within the ETFSI approach with
BSk24~\cite{Pearson_ea18_bsk22-26}, we have recalculated the equation of state for the inner crust. 
Although the corrections to the pressure amount 
to a few percents, the effect  of these corrections for the adiabatic index is much larger: the suppression 
of $\Gamma$ at densities around 0.08~fm$^{-3}$ becomes much less pronounced (previously $\Gamma$ at these densities was underestimated by a factor of two).
Numerical results can be found in the Supplementary Material. 
This corrected equation of state should therefore be preferred for dynamical studies of neutron stars, such 
as the determination of the oscillation modes.

\begin{acknowledgments}
This work benefited from valuable discussions with J.-M. Pearson and M.E. Gusakov to whom the authors are grateful. The work of N.N.S. was 
financially supported by the FWO (Belgium) and the Fonds de la Recherche Scientifique (Belgium) under the Excellence of Science (EOS) programme (project No. 40007501). This work also received funding from the Fonds de la Recherche Scientifique (Belgium) under Grant No. IISN 4.4502.19. 
The work of A.I.C.\ was supported by the Ministry
of Science and Higher Education of the Russian Federation under the state assignment FFUG-2024-0002 of the Ioffe
Institute.
\end{acknowledgments}

\appendix
\section{Functional derivation}
\label{app:functional-derivative}

Let us consider that the energy $E$ is a functional of some density $n(\pmb{r})$. The functional derivative with respect to $n(\pmb{r})$ is defined by 
\beqy
\delta E=\int d^3r\,  \frac{\delta E}{\delta n(\pmb{r})} \delta n(\pmb{r}) \equiv \lim_{\varepsilon\rightarrow 0} \frac{E[n(\pmb{r})+\varepsilon\delta n(\pmb{r})] - E[n(\pmb{r})]}{\varepsilon}=\frac{dE}{d\varepsilon}\biggr\vert_{\varepsilon=0}
\eeqy
where $\delta n(\pmb{r})$ is an \emph{arbitrary} variation of the density that vanishes at the boundary of the integration domain. In this paper, we are only 
interested in semilocal functionals of the kind $\displaystyle E[n(\pmb{r})]=\int d^3r\, \mathcal{E}(\pmb{r})$ with 
$\mathcal{E}(\pmb{r})$ a function of $n(\pmb{r})$ and its derivatives up to the fourth order. The functional derivative is thus given by
\beqy
\delta E=\int d^3r\, \biggl[\frac{\partial \mathcal{E}(\pmb{r})}{\partial n(\pmb{r})}
\delta n(\pmb{r})+\frac{\partial \mathcal{E}(\pmb{r})}{\partial \pmb{\nabla}n(\pmb{r})}\cdot \pmb{\nabla}\delta n(\pmb{r})
+\frac{\partial \mathcal{E}(\pmb{r})}{\partial \nabla^2 n(\pmb{r})} \nabla^2 \delta n(\pmb{r}) \notag \\ 
+\frac{\partial \mathcal{E}(\pmb{r})}{\partial \pmb{\nabla} \nabla^2 n(\pmb{r})} \cdot \pmb{\nabla} \nabla^2 \delta n(\pmb{r})
+\frac{\partial \mathcal{E}(\pmb{r})}{\partial \nabla^4 n(\pmb{r})} \nabla^4 \delta n(\pmb{r})
\biggr]\, .
\eeqy
Integrating by parts, we find 
\beqy
\delta E
=\int d^3r\,  \biggl[\frac{\partial \mathcal{E}(\pmb{r})}{\partial n(\pmb{r})}-\pmb{\nabla}\cdot \frac{\partial \mathcal{E}(\pmb{r})}{\partial \pmb{\nabla}n(\pmb{r})}
+\nabla^2\frac{\partial \mathcal{E}(\pmb{r})}{\partial \nabla^2 n(\pmb{r})} \notag \\ 
-\pmb{\nabla}\cdot  \nabla^2\frac{\partial \mathcal{E}(\pmb{r})}{\partial \pmb{\nabla}\nabla^2 n(\pmb{r})} 
+\nabla^4\frac{\partial \mathcal{E}(\pmb{r})}{\partial \nabla^4 n(\pmb{r})} 
\biggr]\delta n(\pmb{r}) \, .
\eeqy
Since $\delta n(\pmb{r})$ is arbitrary, we obtain for the functional derivative of the energy 
\beqy
\frac{\delta E}{\delta n(\pmb{r})} = \frac{\partial \mathcal{E}(\pmb{r})}{\partial n(\pmb{r})}-\pmb{\nabla}\cdot \frac{\partial \mathcal{E}(\pmb{r})}{\partial \pmb{\nabla}n(\pmb{r})}
+\nabla^2\frac{\partial \mathcal{E}(\pmb{r})}{\partial \nabla^2 n(\pmb{r})}
-\pmb{\nabla}\cdot  \nabla^2\frac{\partial \mathcal{E}(\pmb{r})}{\partial \pmb{\nabla}\nabla^2 n(\pmb{r})} 
+\nabla^4\frac{\partial \mathcal{E}(\pmb{r})}{\partial \nabla^4 n(\pmb{r})} 
\, .
\eeqy
Note that the functional derivative of the energy $E$ with respect to the density $n(\pmb{r})$ has the dimension of energy. 

\section{Energy minimization in the restricted ETF approach}
\label{app:chemical-potential}

In this appendix, the expressions of the chemical potentials $\tilde{\mu}_q$ and $\tilde{\mu}_e$ in the restricted ETF method are derived by minimizing the energy $E_{\rm ETF}$ rather than the grand potential $E_{\rm ETF}-\tilde{\mu}_n N^{(\mathrm{c})}_n-\tilde{\mu}_p N^{(\mathrm{c})}_p-\tilde{\mu}_e N^{(\mathrm{c})}_e$. The numbers $N^{(\mathrm{c})}_q$ of nucleons and $N_e^{(\mathrm{c})}$ of electrons are now fixed \emph{before} minimization (the volume of the cell $V_\mathrm{c}$ is also given). The parameters of the nucleon density distribution $n_q(r)$ are thus related by the requirement 
\beqy\label{eq:constraint-fixed-Nq}
\delta N^{(\mathrm{c})}_q = \frac{\partial N^{(\mathrm{c})}_q}{\partial n_{Bq}}\delta n_{Bq} + \frac{\partial N^{(\mathrm{c})}_q}{\partial \pmb{x_q}}\cdot \delta \pmb{x_q} 
+ \frac{\partial N^{(\mathrm{c})}_q}{\partial n_{\Lambda q}}\delta n_{\Lambda q} = 0\, , 
\eeqy
and similarly for electrons 
\beqy\label{eq:constraint-fixed-Ne}
\delta N^{(\mathrm{c})}_e = \frac{\partial N^{(\mathrm{c})}_e}{\partial n_{e}}\delta n_{e} = 0\, . 
\eeqy

From the integration of Eq.~\eqref{eq:parametrized-profiles1}, we have 
\beqy\label{eq:integrated-Nq}
N^{(\mathrm{c})}_q = n_{B q}V_\mathrm{c}+ n_{\Lambda q}  \int{\rm d}^3\pmb{r}\,f_q(r,\pmb{x_q},R) 
\, ,
\eeqy 
therefore 
\beqy\label{eq:derivative-Nq-nBq}
\frac{\partial N^{(\mathrm{c})}_q}{\partial n_{Bq}} = V_\mathrm{c} \, ,
\eeqy
\beqy \label{eq:derivative-Nq-xq}
\frac{\partial N^{(\mathrm{c})}_q}{\partial \pmb{x_q}} = n_{\Lambda q}  \int{\rm d}^3\pmb{r}\,\frac{\partial f_q(r,\pmb{x_q},R)}{\partial \pmb{x_q}} \, ,
\eeqy 
\beqy \label{eq:derivative-Nq-nLq}
\frac{\partial N^{(\mathrm{c})}_q}{\partial n_{\Lambda q}}=\int{\rm d}^3\pmb{r}\,f_q(r,\pmb{x_q},R)  \, .
\eeqy 
Similarly, 
\beqy\label{eq:derivative-Ne-ne}
\frac{\partial N^{(\mathrm{c})}_e}{\partial n_{e}} = V_\mathrm{c} \, .
\eeqy
It follows from Eqs.~\eqref{eq:constraint-fixed-Ne} and \eqref{eq:derivative-Ne-ne} that the electron number density is fixed and is simply given 
by $n_e=N^{(\mathrm{c})}_e/V_\mathrm{c}$. 

The minimization of the energy can be expressed as  
\beqy
\delta E_{\rm ETF} =  \sum_q\left(\frac{\partial E_{\rm ETF}}{\partial n_{Bq}}\delta n_{Bq} + \frac{\partial E_{\rm ETF}}{\partial \pmb{x_q}}\cdot \delta \pmb{x_q} 
+ \frac{\partial E_{\rm ETF}}{\partial n_{\Lambda q}}\delta n_{\Lambda q} \right)  = 0\, .
\eeqy
Not all variations of the parameters are independent. Eliminating for instance $\delta n_{Bq}$ using Eq.~\eqref{eq:constraint-fixed-Nq} yields 
\beqy 
\sum_q\left(\frac{\partial E_{\rm ETF}}{\partial n_{\Lambda q}}-\frac{1}{V_\mathrm{c}}\frac{\partial E_{\rm ETF}}{\partial n_{Bq}}\frac{\partial N^{(\mathrm{c})}_q}{\partial n_{\Lambda q}}\right)\delta n_{\Lambda q}
+\sum_q\left(\frac{\partial E_{\rm ETF}}{\partial \pmb{x_q}}-\frac{1}{V_\mathrm{c}}\frac{\partial N^{(\mathrm{c})}_q}{\partial \pmb{x_q}}\frac{\partial E_{\rm ETF}}{\partial n_{Bq}}\right)\cdot \delta \pmb{x_q} = 0\, .
\eeqy 
Since $\delta n_{\Lambda q}$ and $\delta \pmb{x_q}$ are arbitrary, we must have 
\beqy 
\frac{\partial E_{\rm ETF}}{\partial n_{\Lambda q}}=\frac{1}{V_\mathrm{c}}\frac{\partial N^{(\mathrm{c})}_q}{\partial n_{\Lambda q}}\frac{\partial E_{\rm ETF}}{\partial n_{Bq}}\, , 
\eeqy 
\beqy 
\frac{\partial E_{\rm ETF}}{\partial \pmb{x_q}}=\frac{1}{V_\mathrm{c}}\frac{\partial N^{(\mathrm{c})}_q}{\partial \pmb{x_q}}\frac{\partial E_{\rm ETF}}{\partial n_{Bq}}\, .
\eeqy
These are the analogs
Euler-Lagrange equations of the variational problem. Once the solutions for $n_{\Lambda q}$ and $\pmb{x_q}$ have been found, the remaining parameters $n_{Bq}$ can be inferred from Eq.~\eqref{eq:integrated-Nq}. 

To obtain the chemical potentials, let us now consider small variations of the particle numbers. From Eq.~\eqref{eq:chemical-potential-general}, the energy change $\delta E_{\rm ETF}$ is given by 
\beqy 
\delta E_{\rm ETF}=\sum_q \tilde{\mu}_q\delta N^{(\mathrm{c})}_q + \tilde{\mu}_e \delta N^{(\mathrm{c})}_e =\sum_q \left( \tilde{\mu}_q \frac{\partial N^{(\mathrm{c})}_q}{\partial n_{Bq}}\delta n_{Bq} + \tilde{\mu}_q\frac{\partial N^{(\mathrm{c})}_q}{\partial \pmb{x_q}}\cdot \delta \pmb{x_q} 
+ \tilde{\mu}_q\frac{\partial N^{(\mathrm{c})}_q}{\partial n_{\Lambda q}}\delta n_{\Lambda q}\right) + \tilde{\mu}_e\frac{\partial N^{(\mathrm{c})}_e}{\partial n_{e}}\delta n_{e}\, ,
\eeqy 
where derivatives must be evaluated for the unperturbed solution. 
From this equation, we thus obtain the identities 
\beqy\label{eq:chemical-potential-identity1} 
\frac{\partial E_{\rm ETF}}{\partial n_{B q}} = \tilde{\mu}_q \frac{\partial N^{(\mathrm{c})}_q}{\partial n_{Bq}} 
\, , 
\eeqy 
\beqy\label{eq:chemical-potential-identity2}  
\frac{\partial E_{\rm ETF}}{\partial \pmb{x_q}} =\tilde{\mu}_q\frac{\partial N^{(\mathrm{c})}_q}{\partial \pmb{x_q}}\, ,
\eeqy 
\beqy\label{eq:chemical-potential-identity3}  
\frac{\partial E_{\rm ETF}}{\partial n_{\Lambda q}} = \tilde{\mu}_q \frac{\partial N^{(\mathrm{c})}_q}{\partial n_{\Lambda q}} 
\, ,
\eeqy
\beqy\label{eq:chemical-potential-identity-elec}  
\frac{\partial E_{\rm ETF}}{\partial n_{e}} = \tilde{\mu}_e \frac{\partial N^{(\mathrm{c})}_e}{\partial n_{e}} \, .
\eeqy 
Using Eq.~\eqref{eq:variation_param-profiles}, variations of the energy can be alternatively written as 
\beqy 
\delta E_{\rm ETF}&=&\sum_q \int {\rm d}^3\pmb{r}\, \frac{\delta E_{\rm ETF}}{\delta n_q(\pmb{r})}\delta n_q(\pmb{r}) + \int {\rm d}^3\pmb{r}\, \frac{\delta E_{\rm ETF}}{\delta n_e(\pmb{r})}\delta n_e(\pmb{r}) \\ 
&=&\sum_q \int {\rm d}^3\pmb{r}\, \frac{\delta E_{\rm ETF}}{\delta n_q(\pmb{r})}\left(\frac{\partial n_q(r)}{\partial n_{Bq}}\delta n_{Bq} + \frac{\partial n_q(r)}{\partial \pmb{x_q}}\cdot \delta \pmb{x_q} 
+ \frac{\partial n_q(r)}{\partial n_{\Lambda q}}\delta n_{\Lambda q}\right) 
+ \int {\rm d}^3\pmb{r}\, \frac{\delta E_{\rm ETF}}{\delta n_e(\pmb{r})}\delta n_e \, , 
\eeqy 
from which we can obtain explicit expressions for the partial derivatives of the energy 
\beqy\label{eq:partial-derivative-energy-nBq} 
\frac{\partial E_{\rm ETF}}{\partial n_{B q}} = \int {\rm d}^3\pmb{r}\, \frac{\delta E_{\rm ETF}}{\delta n_q(\pmb{r})}\frac{\partial n_q(r)}{\partial n_{Bq}} = \int {\rm d}^3\pmb{r}\, \frac{\delta E_{\rm ETF}}{\delta n_q(\pmb{r})}\, ,
\eeqy 
\beqy \label{eq:partial-derivative-energy-xq} 
\frac{\partial E_{\rm ETF}}{\partial \pmb{x_q}} = \int {\rm d}^3\pmb{r}\, \frac{\delta E_{\rm ETF}}{\delta n_q(\pmb{r})}\frac{\partial n_q(r)}{\partial \pmb{x_q}} = n_{\Lambda q} \int {\rm d}^3\pmb{r}\, \frac{\delta E_{\rm ETF}}{\delta n_q(\pmb{r})} \frac{\partial f_q(r,\pmb{x_q},R)}{\partial \pmb{x_q}}\, ,
\eeqy 
\beqy\label{eq:partial-derivative-energy-nLq}  
\frac{\partial E_{\rm ETF}}{\partial n_{\Lambda q}} = \int {\rm d}^3\pmb{r}\, \frac{\delta E_{\rm ETF}}{\delta n_q(\pmb{r})}\frac{\partial n_q(r)}{\partial n_{\Lambda q}} =\int {\rm d}^3\pmb{r}\, \frac{\delta E_{\rm ETF}}{\delta n_q(\pmb{r})} f_q(r,\pmb{x_q},R) \, ,
\eeqy 
\beqy \label{eq:partial-derivative-energy-ne} 
\frac{\partial E_{\rm ETF}}{\partial n_{e}} = \int {\rm d}^3\pmb{r}\, \frac{\delta E_{\rm ETF}}{\delta n_e(\pmb{r})}\frac{\partial n_e(r)}{\partial n_{e}} = \int {\rm d}^3\pmb{r}\, \frac{\delta E_{\rm ETF}}{\delta n_e(\pmb{r})}\, . 
\eeqy 
The derivatives of $n_q(r)$ were calculated from Eq.~\eqref{eq:parametrized-profiles1}. 

Replacing Eqs.~\eqref{eq:derivative-Nq-nBq}-\eqref{eq:derivative-Ne-ne} and \eqref{eq:partial-derivative-energy-nBq}-\eqref{eq:partial-derivative-energy-ne} in Eqs.~\eqref{eq:chemical-potential-identity1}-\eqref{eq:chemical-potential-identity-elec} leads to the same expressions 
\eqref{eq:tilda_mu}, \eqref{eq:chemical-potential-restricted1}, and \eqref{eq:chemical-potential-restricted3} and \eqref{eq:electron-chemical-potential-restricted} obtained from the minimization of the grand potential. 

\section{Chemical potential of nuclei in the outer crust of a neutron star}
\label{app:chemical-potential-outer-crust}

In the outer crust of a neutron star, the `true' chemical potential of nuclei is given within the ETF approach  by 
\beqy\label{eq:true-cluster-chemical-potential}
\mu_X \equiv E_{\rm ETF} - V_\mathrm{c} \, \mathcal{E}_{\rm ETF}(R) \, . 
\eeqy 
Assuming that electrons are uniformly distributed, it can be easily seen from Eq.~\eqref{eq:energy-contributions} that the contributions due to the kinetic energy of an ideal relativistic electron Fermi gas and the electron Coulomb exchange both cancel. Since nucleons are all bound inside clusters, we have $\mathcal{E}_{\rm nuc}(R)=0$. 
Recalling that ${\mathcal E}_{\rm Coul, dir}(R)=0$ due to electric charge neutrality, Eq.~\eqref{eq:true-cluster-chemical-potential} can be written as 
\beqy \label{eq:true-cluster-chemical-potential-outer}
\mu_X = M(A,Z)c^2 + E^{ee}_{\rm Coul,dir} + E^{ep}_{\rm Coul,dir} \, ,
\eeqy 
where $M(A,Z)c^2\equiv E_{\rm nuc} +  E^{pp}_{\rm Coul,dir}+E^{pp}_{\rm Coul,exch}$ represents the nuclear mass in vacuum, and we have decomposed the direct and exchange Coulomb energies 
into contributions due to proton-proton (pp), electron-electron (ee), and electron-proton (ep) Coulomb interactions. 

Using Eqs.~\eqref{eq:direct-Coulomb-energy} and \eqref{eq:Coulomb-potential}, the electron-electron part can be readily obtained  
\beqy\label{eq:direct-Coulomb-energy-ee}
E^{ee}_{\rm Coul,dir}=\frac{16}{15}\pi^2 e^2 n_e^2 R^5\, . 
\eeqy 
The electron-proton part is given by 
\beqy\label{eq:direct-Coulomb-energy-ep} 
E^{ep}_{\rm Coul,dir}=-8\pi^2e^2 n_e \int_0^R {\rm d}r\, r^2 \int_0^R {\rm d}r'\, r'^2 n_p(r') \frac{r+r'-\vert r-r'\vert}{rr'}\, .
\eeqy 
Considering point-like nuclei, the proton density distribution is approximated by 
\beqy 
n_p(r)\approx \dfrac{Z}{4\pi r^2}\delta(r) = \frac{R^3}{3r^2}n_e \delta(r) \, ,
\eeqy 
where $\delta(r)$ denotes the Dirac-delta distribution and we have made use of the relation 
\beqy\label{eq:np-vs-Z}
\bar n_p=n_e=\frac{3 Z}{4\pi R^3} \, .
\eeqy
The integral in Eq.~\eqref{eq:direct-Coulomb-energy-ep} can then be calculated analytically: 
\beqy\label{eq:direct-Coulomb-energy-ep-final} 
E^{ep}_{\rm Coul,dir}=-\frac{8}{3}\pi^2e^2 n_e^2 R^5\, .
\eeqy
Substituting Eqs.~\eqref{eq:direct-Coulomb-energy-ee} and \eqref{eq:direct-Coulomb-energy-ep-final} into \eqref{eq:true-cluster-chemical-potential-outer} leads to \eqref{eq:chemical-potential-nuclei}.

\section{Pressure formulas}

\subsection{Hartree-Fock theory}
\label{app:pressure-HF}

Adopting the spherical Wigner-Seitz approximation, the pressure formula~\eqref{eq:pressure-variations} in the HF theory reads 
\beqy\label{eq:pressure-variations-HF}
P_{\rm HF}=-\mathcal{E}_{\rm HF}(R)- \frac{1}{R^2}\int_0^R{\rm d}r\, r^2 
\sum_q\biggl[U_q(r)\frac{\partial n_q(r)}{\partial R}
+\frac{\hbar^2}{2 m^\oplus_q(r)}\frac{\partial \tau_q(r)}{\partial R}
+ W_q(r) \frac{\partial J_q(r)}{\partial R}\biggr]\notag \\ 
- \frac{1}{R^2}\int_0^R{\rm d}r\, r^2 U_e(r)\frac{\partial n_e(r)}{\partial R}\, .
\eeqy

Let us examine the first term in the integral over nucleon fields: 
\beqy
\int_0^R{\rm d}r\, r^2 
U_q(r)\frac{\partial n_q(r)}{\partial R}
=\frac{1}{2\pi}\sum_{\alpha(q)}g^{(q)}_\alpha  \int_0^R{\rm d}r\, r^2 
U_q(r) \mathcal{R}^{(q)}_{\alpha}(r)\frac{\partial \mathcal{R}^{(q)}_{\alpha}}{\partial R}   \, .
\eeqy
Now the second term: 
\beqy
&&\int_0^R{\rm d}r\, r^2 
\frac{\hbar^2}{2 m_q^\oplus(r)}\frac{\partial \tau_q(r)}{\partial R} \notag \\ 
&=& \frac{1}{2\pi}\sum_{\alpha(q)}g^{(q)}_\alpha\int_0^R{\rm d}r\, r^2 
\frac{\hbar^2}{2 m_q^\oplus(r)} \Biggl[\frac{d\mathcal{R}^{(q)}_{\alpha}}{dr}\frac{d}{dr}\frac{\partial \mathcal{R}^{(q)}_{\alpha}}{\partial R}+\frac{\ell_\alpha(\ell_\alpha+1)}{r^2} \mathcal{R}^{(q)}_{\alpha}(r)\frac{\partial  \mathcal{R}^{(q)}_{\alpha}}{\partial R} \Biggr] \notag \\ 
&=&\frac{1}{2\pi}\sum_{\alpha(q)}g^{(q)}_\alpha \int_0^R{\rm d}r\,\Biggl[ -\frac{d}{dr} r^2 
\frac{\hbar^2}{2 m_q^\oplus(r)} \frac{d\mathcal{R}^{(q)}_{\alpha}}{dr} \Biggr]\frac{\partial \mathcal{R}^{(q)}_{\alpha}}{\partial R} \notag \\
&&+\frac{1}{2\pi}\sum_{\alpha(q)}g^{(q)}_\alpha \int_0^R{\rm d}r\, r^2 \frac{\hbar^2}{2 m_q^\oplus(r)} \frac{\ell_\alpha(\ell_\alpha+1)}{r^2} \mathcal{R}^{(q)}_{\alpha}(r)\frac{\partial  \mathcal{R}^{(q)}_{\alpha}}{\partial R} \notag \\ 
&=&\frac{1}{2\pi}\sum_{\alpha(q)}g^{(q)}_\alpha \int_0^R{\rm d}r\,\Biggl[ -\frac{d}{dr} r^2 
\frac{\hbar^2}{2 m_q^\oplus(r)} \frac{d}{dr} 
+\frac{\hbar^2}{2 m_q^\oplus(r)} \ell_\alpha(\ell_\alpha+1) \Biggr]\mathcal{R}^{(q)}_{\alpha}(r)\frac{\partial  \mathcal{R}^{(q)}_{\alpha}}{\partial R} \, ,
\eeqy
where we used integration by parts.
Then the last term: 
\beqy
&&\int_0^R{\rm d}r\, r^2 W_q(r)  \frac{\partial J_q(r)}{\partial R} \notag \\ 
&=& \frac{1}{2\pi}  \sum_{\alpha(q)}g^{(q)}_\alpha  \left[j_\alpha(j_\alpha+1)-\ell_\alpha(\ell_\alpha+1)-\frac{3}{4}\right] \int_0^R{\rm d}r\, r W_q(r) \mathcal{R}^{(q)}_{\alpha} (r)\frac{\partial  \mathcal{R}^{(q)}_{\alpha}}{\partial R} \, .
\eeqy 
Collecting all terms and recalling that the single-particle wave functions must be those solving 
the HF equation~\eqref{eq:HF-WS}, we find that the pressure can be written as 
\beqy\label{eq:pressure-HF}
P_{\rm HF}=-\mathcal{E}_{\rm HF}(R)
-\frac{1}{2\pi R^2}\sum_q\sum_{\alpha(q)}g^{(q)}_\alpha \epsilon^{(q)}_\alpha \int_0^R{\rm d}r\, r^2
 \mathcal{R}^{(q)}_{\alpha} (r) \frac{\partial  \mathcal{R}^{(q)}_{\alpha}(r)}{\partial R}
\notag \\ - \frac{1}{R^2}\int_0^R{\rm d}r\, r^2 U_e(r)\frac{\partial n_e(r)}{\partial R}\, .
\eeqy
Now, the number of nucleons in the cell is given by 
\beqy 
N^{(\mathrm{c})}_q=4\pi \int_0^R{\rm d}r\, r^2 n_q(r) =  \sum_{\alpha(q)}g^{(q)}_\alpha  \int_0^R{\rm d}r\, r^2 \mathcal{R}^{(q)}_{\alpha}(r)^2 \, .
\eeqy 
Since the derivative with respect to $R$ must be evaluated keeping $N^{(\mathrm{c})}_q$ fixed, we have 
\beqy 
0= \sum_{\alpha(q)}g^{(q)}_\alpha \Biggl[ R^2 \mathcal{R}^{(q)}_{\alpha}(R)^2 + 2\int_0^R{\rm d}r\, r^2 \mathcal{R}^{(q)}_{\alpha}(r)  \frac{\partial  \mathcal{R}^{(q)}_{\alpha}(r)}{\partial R} \Biggr]\, .
\eeqy 
This equation must hold irrespective of the occupation factors. Therefore, we obtain the following identity: 
\beqy 
\int_0^R{\rm d}r\, r^2 \mathcal{R}^{(q)}_{\alpha}(r)  \frac{\partial  \mathcal{R}^{(q)}_{\alpha}(r)}{\partial R} =-\frac{1}{2} R^2 \mathcal{R}^{(q)}_{\alpha}(R)^2\, .
\eeqy 
Substituting into Eq.~\eqref{eq:pressure-HF} leads to 
\beqy\label{eq:pressure-HF-final}
P_{\rm HF}=-\mathcal{E}_{\rm HF}(R)+
\frac{1}{4\pi }\sum_q\sum_{\alpha(q)}g^{(q)}_\alpha \epsilon^{(q)}_\alpha \mathcal{R}^{(q)}_{\alpha}(R)^2  - \frac{1}{R^2}\int_0^R{\rm d}r\, r^2 U_e(r)\frac{\partial n_e(r)}{\partial R}\, .
\eeqy

Neglecting electron charge screening effects by considering $n_e(r)\approx n_e$, the pressure can be expressed as Eq.~\eqref{eq:pressure-HF-approx}.

\subsection{Extended Thomas-Fermi approach with parametrized profiles}
\label{app:pressure-ETF-param}

Quite generally, expressing the profiles as in Eq.~\eqref{eq:parametrized-profiles1}, we have 
\beqy\label{eq:variations-profiles}
\frac{\partial n_q(r)}{\partial R} = \frac{\partial n_q(r)}{\partial n_{Bq}}\frac{\partial n_{Bq}}{\partial R} + \frac{\partial n_q(r)}{\partial \pmb{x_{q}}}\cdot \frac{\partial \pmb{x_{q}}}{\partial R}+\frac{\partial n_q(r)}{\partial n_{\Lambda q}}\frac{\partial n_{\Lambda q}}{\partial R}+n_{\Lambda q}\frac{\partial f_q(r)}{\partial R}\biggr\vert_{\pmb{x_q}}\, ,
\eeqy
where the last term arises from the explicit dependence of the parametrized distributions on $R$. Let us analyze separately the contribution of each term to the pressure, given in general by Eq.~\eqref{eq:pressure-variations-ETF}. Substituting 
the first term of Eq.~\eqref{eq:variations-profiles} leads to the pressure contribution 
\beqy
- \frac{1}{R^2}\sum_q\frac{\partial n_{Bq}}{\partial R}  \int_0^R{\rm d}r\, r^2 
\frac{\delta E_{\rm ETF}}{\delta n_q(r)} \frac{\partial n_q(r)}{\partial n_{Bq}}\, .
\eeqy
This integral can be easily evaluated using Eq.~\eqref{eq:delta_E_ETF_param1}: 
\beqy
- \frac{1}{R^2}\sum_q \tilde{\mu}_q \frac{\partial n_{Bq}}{\partial R} \int_0^R{\rm d}r\, r^2 \,  \frac{\partial n_q(r)}{\partial n_{Bq}} \, .
\eeqy
The next two terms can be treated in the same way using Eqs.~\eqref{eq:delta_E_ETF_param2} and \eqref{eq:delta_E_ETF_param4}, respectively. This leads to 
\begin{eqnarray}
&-& \frac{1}{R^2}\sum_q \tilde{\mu}_q  \int_0^R{\rm d}r\, r^2 \, \biggl[ \frac{\partial n_q(r)}{\partial n_{Bq}}\frac{\partial n_{Bq}}{\partial R} 
+\frac{\partial n_q(r)}{\partial \pmb{x_{q}}}\cdot \frac{\partial \pmb{x_{q}}}{\partial R}
+ \frac{\partial n_q(r)}{\partial n_{\Lambda q}}\frac{\partial n_{\Lambda q}}{\partial R} \biggr]  \nonumber
\\
&=&- \frac{1}{R^2}\sum_q \tilde{\mu}_q  \int_0^R{\rm d}r\, r^2 \, 
\left[ \frac{\partial n_q(r)}{\partial R} 
-n_{\Lambda q}\frac{\partial f_q(r)}{\partial R}\biggr\vert_{\pmb{x_q}}\right]
=\sum_q  \tilde{\mu}_q \left[ n_q(R)
+\frac{n_{\Lambda q}}{R^2} \int_0^R{\rm d}r\, r^2 \, 
\frac{\partial f_q(r)}{\partial R}\biggr\vert_{\pmb{x_q}}\right]
\, .
\end{eqnarray}
Here Eqs.~\eqref{eq:variations-profiles} and \eqref{B7a} were applied in the first and the second equality, respectively. 
Adding the pressure contribution corresponding to the last term of Eq.~\eqref{eq:variations-profiles}, we get
\beqy
\sum_q  \tilde{\mu}_q n_q(R) 
+\sum_q   \frac{1}{R^2}\sum_q n_{\Lambda q} \int_0^R{\rm d}r\, r^2 \ \biggl[\tilde{\mu}_q-\frac{\delta E_{\rm ETF}}{\delta n_q(r)}\biggr]\frac{\partial f_q(r)}{\partial R}\biggr\vert_{\pmb{x_q}} \, .
\eeqy
Let us stress that $\delta E_{\rm ETF}/\delta n_q(r)$ here must be evaluated using the parametrized profiles and thus cannot be replaced by $\mu_q$ because the parametrized profiles do not satisfy the Euler-Lagrange equations~\eqref{eq:Euler-Lagrange}. 

Likewise, the assumption of a uniform electron gas amounts in parametrizing the electron density $n_e(r)$ with a background term $n_e$. Therefore, the electron chemical potential is not exactly given by Eq.~\eqref{eq:Euler-Lagrange-electrons}, as was previously assumed in Ref.~\cite{Pearson+12} and should be evaluated from Eq.~\eqref{eq:electron-chemical-potential-restricted}. 
Collecting all the terms, the pressure can be expressed as 
\beqy\label{eq:pressure-param}
\tilde{P}_{\rm ETF}=-\mathcal{E}_{\rm ETF}(R)+\tilde{\mu}_e n_e+\sum_q \tilde{\mu}_q n_{q}(R) +\frac{1}{R^2}\sum_q n_{\Lambda q} \int_0^R{\rm d}r\, r^2 \ \biggl[\tilde{\mu}_q-\frac{\delta E_{\rm ETF}}{\delta n_q(r)}\biggr]\frac{\partial f_q(r)}{\partial R}\biggr\vert_{\pmb{x_q}} \, .
\eeqy

The use of $\tilde{\mu}_e$ instead of the chemical potential of a homogeneous electron gas allows to recover the previously missing lattice pressure in the formula obtained in Ref.~\cite{Pearson+12}. Comparing Eqs.~\eqref{eq:electron-chemical-potential-restricted} and ~\eqref{eq:mue_hom}, it can be seen that $\tilde{\mu}_e$ contains an additional term given by 
\beqy 
- \frac{3 e}{R^3}\int_0^R {\rm d}r\, r^2 U_{\rm Coul}(r)\, .
\eeqy 
Using Eq.~\eqref{eq:Coulomb-potential}, the 
term $\tilde{\mu}_e n_e$ in Eq.~\eqref{eq:pressure-param} therefore contains the additional pressure contribution
\beqy \label{eq:lattice-pressure-traditional}
P_{\rm Coul,dir}=-\frac{3}{10}\frac{N^{(\mathrm{c})}_e e^2}{R} n_e \left(1-\frac{5}{3}\frac{\langle r^2\rangle}{R^2}\right)\, ,
\eeqy 
where we have made use of the relation 
\beqy\label{eq:ne-vs-Z}
n_e=\frac{3 N^{(\mathrm{c})}_e}{4\pi R^3} \, ,
\eeqy
and we have introduced the mean square radius of the proton distribution defined by Eq.~\eqref{eq:mean-square-proton-radius}. 

Equation~\eqref{eq:lattice-pressure-traditional} coincides with the lattice pressure obtained from the differentiation of Eq.~(5.6) of Ref.~\cite{BBP1971}. 
The formula~\eqref{eq:pressure-param} therefore ensures a correct treatment of Coulomb interactions throughout the crust. 
Equation~\eqref{eq:lattice-pressure-traditional} can be equivalently written as Eq.~\eqref{eq:lattice-pressure} using Eq.~\eqref{eq:ne-vs-Z}.

\bibliography{references}

\end{document}